\documentclass[fleqn, usenatbib]{mnras} 

\usepackage[T1]{fontenc}
\usepackage{ae, aecompl}

\usepackage{graphicx}        
\usepackage{amsmath}         
\usepackage{amssymb}         
\usepackage{booktabs}        
\usepackage{scalerel}        
\usepackage{xcolor}          
\usepackage{upquote}         
\usepackage{subcaption}      
\usepackage{multirow}        
\usepackage{pdflscape}       
\usepackage{microtype}       
\usepackage{pgffor}          
\usepackage{siunitx}         
\usepackage{crossreftools}   
\usepackage{orcidlink}       

\usepackage{enumitem}        
\setlist[enumerate]{style=standard,
                    align=left,
                    labelindent=1em,
                    labelwidth=!,
                    leftmargin=*,
                    itemindent=0em}

\definecolor{key1}{RGB}{122, 5, 137}
\definecolor{key2}{RGB}{34, 19, 163}

\usepackage{listings} 
\newcommand*{\dpi}{600} 

\newcommand*{\orcid}[1]{\;\!\orcidlink{#1}\;\!}

\newcommand*{\round}[2][3]{ \num[round-mode=places, round-precision=#1]{#2} }

\newcommand*{\units}[1]{\scalebox{0.8}{(#1)}}

\newcommand*{\Sun}{\protect\scalebox{0.7}{$\odot$}}

\newcommand*{\Range}[2]{[#1\,\text{,}\,#2]}
\newcommand*{\RangeNI}[2]{[#1\,\text{,}\,#2)}
\newcommand*{\pd}[1]{\times\!10^{#1}}

\newcommand*{\SM}[1]{{\scaleto{\rm #1}{3.5pt}}}

\newcommand*{\var}[1]{\mbox{\footnotesize$(#1)$}}

\newcommand*{\svar}[1]{\!\mbox{\footnotesize$(#1)$}}

\newcommand*{\BPRP}{{G_{\rm BP}\!-\!G_{\rm RP}}}

\newcommand*{\nbody}{\hbox{\textit{N}\!\!\:-body}}
\newcommand*{\stube}{\hbox{S\hspace{0.075em}-\hspace{-0.025em}Tube}}

\newcommand*{\GtrSim}{\smallrel\gtrsim}
\newcommand*{\LessSim}{\smallrel\lesssim}
\newcommand*{\GtrApprox}{\smallrel\gtrapprox}

\newcommand*{\Approx}{\smallrel\sim}

\makeatletter
\newcommand*{\smallrel}[2][.8]{%
  \mathrel{\mathpalette{\smallrel@{#1}}{#2}}%
}
\newcommand*{\smallrel@}[3]{%
  \sbox0{$#2\vcenter{}$}%
  \dimen@=\ht0 %
  \raise\dimen@\hbox{%
    \scalebox{#1}{%
      \raise-\dimen@\hbox{$#2#3\m@th$}%
    }%
  }%
}

\DeclareMathOperator{\logN}{logN}

\DeclareMathOperator{\mean}{mean}
\DeclareMathOperator{\std}{std}

\DeclareMathOperator{\modulus}{mod}
\DeclareMathOperator{\reminder}{rem}
\DeclareMathOperator{\normalised}{norm}

\definecolor{mycolour}{HTML}{ab1f20}

\newcommand*{\TM}{\protect\scalebox{0.65}{TM}}

\newcommand*{\GC}{\protect\scalebox{0.50}{\:\!GC}}
\newcommand*{\D}{\protect\scalebox{0.50}{\:\!D}}
\newcommand*{\M}{\protect\scalebox{0.50}{\:\!M}}
\newcommand*{\R}{\protect\scalebox{0.50}{\:\!R}}
\newcommand*{\RS}{\protect\scalebox{0.50}{\:\!RS}}
\newcommand*{\SD}{\protect\scalebox{0.50}{\:\!SD}}
\newcommand*{\St}{\protect\scalebox{0.50}{\:\!S}}
\newcommand*{\Z}{\:\!\protect\scalebox{0.61}{$0$}}
\newcommand*{\T}{\:\!\protect\scalebox{0.60}{$-T$}}
\newcommand*{\Lim}{\protect\scalebox{0.50}{\:\!L}}
\newcommand*{\Inf}{\protect\scalebox{0.50}{\:\!Inf}}
\newcommand*{\Sup}{\protect\scalebox{0.50}{\:\!Sup}}

\newcommand{\s}[1]{
  \foreach \n in {1,...,#1}{\!}
}

\makeatletter
\newcommand{\optionaldesc}[2]{%
  \phantomsection
  #1\protected@edef\@currentlabel{#1}\label{#2}%
}
\makeatother

\title[M68 stream angle-actions]{Modelling the M68 stellar stream with realistic mass loss and frequency distributions in angle-action coordinates}
\author[C. G. Palau, W. Wang, J. Han]{
Carles G. Palau\orcid{0000-0002-7583-534X}$^{1,2}$\thanks{E-mail: cgpalau@sjtu.edu.cn},
Wenting Wang\orcid{0000-0002-5762-7571}$^{1,2}$, 
Jiaxin Han\orcid{0000-0002-8010-6715}$^{1,2}$\thanks{E-mail: jiaxin.han@sjtu.edu.cn}
\\
$^{1}$Department of Astronomy, Shanghai Jiao Tong University, 800 Dongchuan Road, Shanghai 200240, China\\
$^{2}$Shanghai Key Laboratory for Particle Physics and Cosmology, 800 Dongchuan Road, Shanghai 200240, China\\
}

\date{Accepted XXX. Received YYY; in original form ZZZ}

\pubyear{2025}

\begin{document}
\label{firstpage}
\pagerange{\pageref{firstpage}--\pageref{lastpage}}
\maketitle

\begin{abstract}
We develop a new method for simulating stellar streams generated by globular clusters using angle-action coordinates. This method reproduces the variable mass loss and variable frequency of the stripped stars caused by the changing tidal forces acting on the cluster as it moves along an eccentric orbit. The model incorporates realistic distributions for the stripping angle and frequency of the stream stars both along and perpendicular to the stream. The stream is simulated by generating random samples of stripped stars and integrating them forward in time in angle-frequency space. Once the free parameters are calibrated, this method can be used to simulate the internal structure of stellar streams more quickly than \nbody\ simulations, while achieving a similar level of accuracy. We use this model to study the surface density of the stellar stream produced by the globular cluster M68 (NGC~4590). We select $291$ stars from the \textit{Gaia}-DR3 catalogue along the observable section that are likely to be members of the stream. We find that the width of the stream is too large to be explained by stars stripped from the cluster alone. We simulate the stream using the present method and include the \textit{Gaia} selection function and observational errors, and the process of separating the stream stars from the foreground. By comparing these results with the observed data, we estimate the age of the stream, or equivalently the cluster accretion time, to be $\round[2]{3.0397}_{-\round[2]{0.28704979}}^{+\round[2]{5.63203224}}$~Gyr, and the mass loss of the cluster to be $0.496\pm0.030$~M$_{\Sun}$~Myr$^{-1}$~arm$^{-1}$.
\end{abstract}


\defcitealias{2008gady.book.....B}{BT08}
\defcitealias{2025MNRAS.539.2718P}{P25}



\section{Introduction}\label{introduction}

Stellar streams are formed by stars that have been stripped from a progenitor cluster by tidal forces caused by the host galaxy. Recently, thanks to the data provided by the \textit{Gaia} mission \citep{2016AandA...595A...1G}, many streams have been discovered in the Solar neighborhood \citep{2025NewAR.10001713B}. These streams can provide valuable information about the history and evolution of the Milky Way \cite[e.g.][]{2024NewAR..9901706D} and its characteristics, especially its potential \cite[e.g.][]{2025NewAR.10001721H}. In addition, stellar streams can be used to constrain the properties of the Galactic satellites such as the Large Magellanic Cloud (LMC) \citep[e.g.][]{2023MNRAS.521.4936K} or the spatial and mass distribution of the sub-halos predicted by cosmological simulations \citep[e.g.][]{2017ARA&A..55..343B}.

Stellar streams can also be used to study the formation and evolution of their progenitors, as well as to estimate their current properties. For this purpose, the streams generated by globular clusters offer a unique opportunity because they are dynamically cold, the gas within the stream is expected to be negligible, and in general, they present a simpler structure than the large streams generated by dwarf galaxies such as Sagittarius \citep[e.g.][]{1994Natur.370..194I}. The most promising streams are long, close to the Sun, and have no significant perturbations. The best examples are the streams of M68 \citep{2019MNRAS.488.1535P}, NGC~3201 \citep{2021MNRAS.504.2727P} and Palomar~5 \citep[e.g.][]{2001ApJ...548L.165O}. In such cases, we can use the length and density of the stream to estimate the accretion time and to describe the process of mass loss that the cluster has undergone during its evolution within the potential of the Milky Way.

For a cluster following a general eccentric orbit in the halo of its host galaxy, the density of its stream depends on several factors. Firstly, the shape of the stream changes along the orbit. Its curvature is maximum at the apocentres and minimum at the pericentres. Similarly, the volume of the stream is compressed and expanded at these locations, respectively \citep[e.g.][]{2015MNRAS.446.3100H}. In addition, the density of the stream depends on the mass loss of the progenitor cluster. This is maximal at the pericentres, where the tidal perturbations are strongest. This implies that the stripped stars form clumps that are generated at regular intervals along the cluster orbit \citep[e.g.][]{2004AJ....127.2753D}. The mass loss is also a function of the mass of the progenitor and its density distribution, which evolve as stars are lost along the orbit. Another significant factor is the relative motion of the stream stars with respect to the cluster orbit. Due to small variations in angular phase between the stream stars, their relative motion create over- and under-densities within the stellar stream. This phenomenon is known as epicyclic density variations \citep[e.g.][]{2009MNRAS.392..969J, 2010MNRAS.401..105K, 2012MNRAS.420.2700K}. Finally, the relative velocity of the stream stars with respect to the progenitor makes the stream longer and less dense over time.

Boundary cases allow us to minimise or eliminate some of these factors, making it easier to study the remaining ones. For example, in a circular orbit, the curvature and compression of the stream along the orbit as well as the tidal forces acting on the cluster are constant. In this case, we can isolate the variations in mass loss caused by the evolution of the mass and density profile of the cluster, the epicyclic density variations, and the growth of the stream length over time. A more general approach is to describe the stream in angle-action coordinates \citep[][hereafter \citetalias{2008gady.book.....B}]{2008gady.book.....B}. In this case, the stream has a rectilinear shape \citep{2011MNRAS.413.1852E}. This greatly simplifies its analysis by eliminating the geometry variations along the orbit that are present in other coordinate systems. In addition, by rotating the angle-action coordinates to align them with the principal axes of the stream, the projection distortions are minimised and each coordinate can be studied independently. Furthermore, since the stars follow uniform rectilinear motions, the stream does not present epicyclic density variations. Consequently, the density of the stream depends on the mass loss of the cluster, the frequency distribution of the stream stars, and the time since its inception.

Previous studies have modelled stellar streams using angle-action coordinates. For example, \citet{2014MNRAS.443..423S} and \citet{2014ApJ...795...95B} modelled a stellar stream under the assumption of uniform mass loss from the cluster. Similarly, they assumed that the frequency distribution of the stripped stars is independent of the stripping time. These assumptions are reasonable, given that these models were created to constrain the Galactic potential. In such cases, it is unnecessary to accurately model the details of the stream surface density, since only its large-scale structure needs to be considered. Conversely, \citet{2015MNRAS.452..301F} included a variable mass loss from the progenitor cluster. This variation is caused by the changing tidal forces acting on the cluster as it moves along an eccentric orbit. However, the stream was not fully modelled in angle-action space. Building upon these studies, we present a new methodology for modelling stellar streams in angle-action coordinates. This takes into account both the variable mass loss of the progenitor cluster and the variable frequency distributions of the stripped stars. Such improvements are generally necessary to enable the model to be compared with real stellar streams.

In this paper, we use this methodology to study the stream density of the globular cluster M68 (NGC~4590). We use the methodology, notation, and the \nbody\ simulation introduced in \citep[][hereafter \citetalias{2025MNRAS.539.2718P}]{2025MNRAS.539.2718P}. In this simulation, the globular cluster is modelled using a realistic stellar population that follows an isochrone generated by PARSEC/COLIBRI (\S2.1, \citetalias{2025MNRAS.539.2718P}). This population is then distributed in phase-space according to a generalised King model (\S2.2, \citetalias{2025MNRAS.539.2718P}). The orbit of the cluster is integrated backwards during $T=1.5$~Gyr from the current position of the cluster within an axisymmetric mass model of the Milky Way (\S2.4, \citetalias{2025MNRAS.539.2718P}). This model consists of a spherical bulge, a Miyamoto-Nagai disc, and a dark halo modelled with a Navarro, Frenk \& White (NFW) density profile. The stars are located at the final position of the inverse integration and then integrated forward using the \nbody\ code \textsc{PeTar} (\S2.3, \citetalias{2025MNRAS.539.2718P}). The cluster follows an eccentric regular \stube\ orbit and passes through three pericentre passages (\S2.5, \citetalias{2025MNRAS.539.2718P}).

In this paper we denote the action-angle coordinates by $(\theta_i, J_i)$ and the frequencies by $\varOmega_i$, where $i=\{r,\phi,z\}$. We compute their value using the Averaged Generating Function (AvGF) method implemented in the Python package \texttt{galpy} (\S3, \citetalias{2025MNRAS.539.2718P}). When the coordinates and frequencies are given relative to the centre of the globular cluster, they are denoted by $(\Delta\theta_i, \Delta J_i)$ and $\Delta \varOmega_i$. We apply a rotation in phase-space to align angles, actions and frequencies with the principal axes of the stream \citep{1999MNRAS.307..877T}. The rotation is described in Appendix~\ref{App1}. The rotated coordinates are denoted by $(\Delta\bar{\theta}_j, \Delta\bar{J}_j)$ and the frequencies by $\Delta\bar{\varOmega}_j$, where $j=1$ corresponds to the direction aligned with the principal axis of the stream and $j=\{2,3\}$ correspond to the perpendicular directions.

In Section~\ref{mass_loss}, we describe the mass loss of the globular cluster along its orbit and present the internal structure of the stellar stream based on the \nbody\ simulation. In Section~\ref{model}, we model the stellar stream in angle-action coordinates. The first application of this model is to determine the accretion time and the mass loss of the M68 globular cluster. To achieve this, we compare the surface density of the stream predicted by the model to the observed stream star data. In Section~\ref{GDR3_selection} we present our selection of stream star candidates from the \textit{Gaia} catalogue. Section~\ref{mock_selection} describes how we generate mock observations from the stream model. These mocks include the \textit{Gaia} selection function, observational uncertainties, and the process of separation of stream stars from the foreground that we use to identify the real stream stars. In Section~\ref{estimations} we compare the observed surface density with the mock distribution and estimate the accretion time of the cluster and the mass loss with a Likelihood technique. Finally, we present our conclusion in Section~\ref{conclusion}.

\section{Mass loss along an eccentric orbit}\label{mass_loss}

The mass loss of the progenitor cluster is proportional to the tidal forces exerted by the host potential (\S7.5.6, \citetalias{2008gady.book.....B}). For a cluster with an eccentric orbit, the tidal forces increase to a relative maximum at the pericentres. Thus, the cluster undergoes a tidal shock at each pericentre passage and the number of stars stripped peaks. The tidal forces generated by the disc crossings are stronger, but they act for a short time and only result in an increase in the velocity dispersion of the cluster (\S8.2.2.f, \citetalias{2008gady.book.....B}). For a spherical Galactocentric radius $r\GtrSim R_{\Sun}$, we consider this effect negligible \citep{1997MNRAS.289..898V}.

\begin{figure*}
\includegraphics[width=1.0\textwidth]{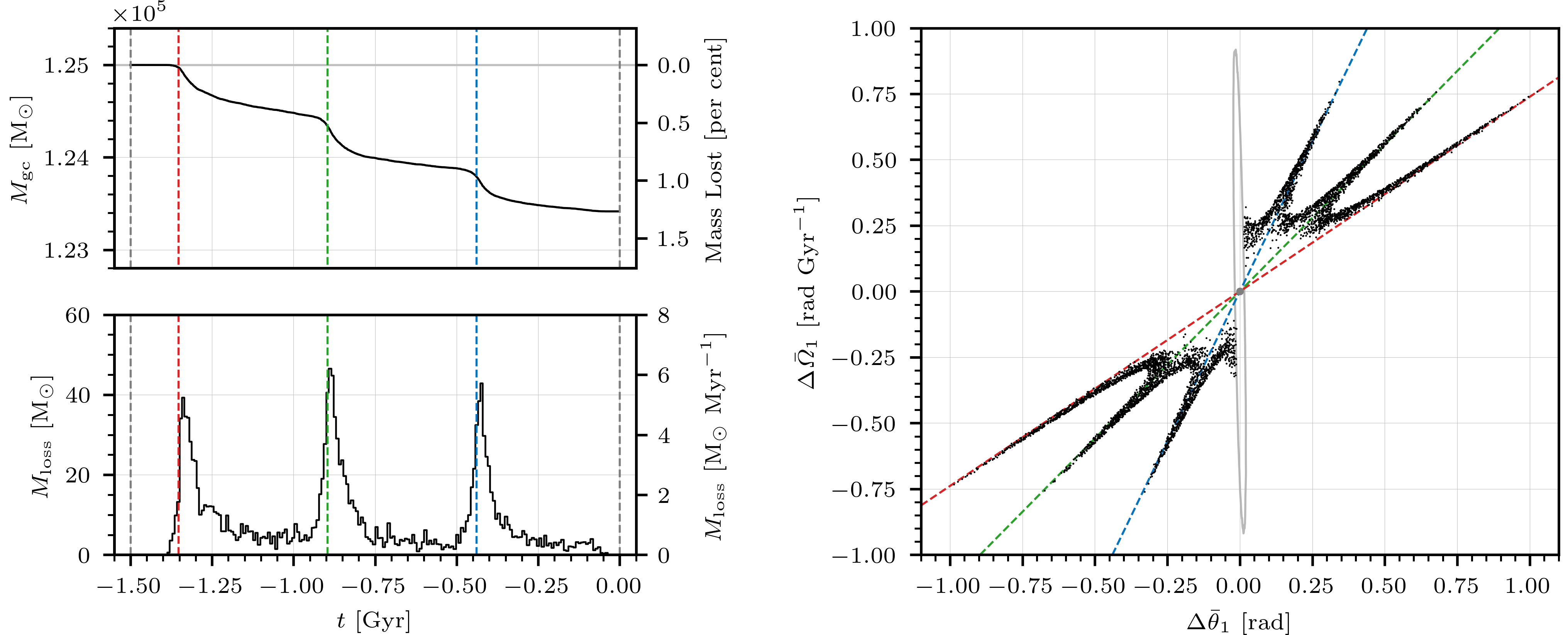}
\caption{\textit{Top left:} Mass of the M68 globular cluster $M_{\rm gc}$ as a function of the simulation time $t$. The coloured dashed lines mark the time of the first (red), second (green), and third (blue) pericentre passages, while the grey dashed lines mark the start and end of the simulation. The solid grey line marks the initial mass of the cluster $M_{\rm ini}$. \textit{Bottom left:} Histogram of the mass loss as a function of the simulation time in bins of 7.5~Myr. \textit{Right:} Relative angle and frequency along the principal axis of the stream \{$\Delta\bar{\theta}_1, \Delta\bar{\varOmega}_1$\}. The large grey dot marks the position of the cluster and the grey contour line the area containing 68 per cent of the cluster stars. The coloured dashed lines mark the frequency of test particles ejected at the pericentre passages with frequencies distributed uniformly as a function of the angle.}
\label{m_loss}
\end{figure*}

In the top left panel of Figure~\ref{m_loss} we show the mass of M68 $M_{\rm gc}$ as a function of the simulation time $t$, where $t=-1.5$~Gyr corresponds to the start of the simulation, and $t=0$ corresponds to the current time. In the lower left panel we show the mass loss of the progenitor cluster $M_{\rm loss}$ in bins of $7.5$~Myr. The mass loss as a function of $t$ is estimated from the stripping time $t_{\rm s}$ of each stream star (Eq.~8, \citetalias{2025MNRAS.539.2718P}). The red, blue, and green vertical dashed lines mark the times of the three pericentre passages that the cluster has undergone since the start of the simulation $t_{\rm peri}\equiv(-1353.59,\, -896.13,\, -439.64)$~Myr. The mass loss begins when the cluster approaches the first pericentre and is heated by the tidal shock. Previously, the number of stripped stars is zero. Between the pericentres, the mass loss is approximately constant at $\Approx 0.5$~M$_{\Sun}$~Myr$^{-1}$, and it increases to $\Approx 5.5$~M$_{\Sun}$~Myr$^{-1}$ at the peaks. The mass loss over the last $50$~Myr is not accounted for, as the stripped stars are close to the cluster and are not classified as stream stars (App.~D, \citetalias{2025MNRAS.539.2718P}). During the $1.5$~Gyr of the simulated orbit, the cluster loses about $0.34$ per cent of its initial mass ($M_{\rm ini}=1.25\pd{5}$~M$_{\Sun}$) at each pericentre passage, and about $1.27$ per cent in total.

In general, for a cluster with an eccentric orbit, the stripped stars at each pericentre generate a pair of symmetric stellar streams \citep[e.g.][]{2004AJ....127.2753D}. Each pair expands along the principal axis of the stream, one at a higher frequency (leading arm) and the other at a lower frequency (trailing arm) with respect to the progenitor cluster. In this way, the tidal stream is formed by a series of small internal streams. In cartesian Galactocentric coordinates, the internal streams appear separated from each other when the progenitor is close to the apocentre, and they overlap when the cluster is close to the pericentre \citep{2015MNRAS.446.3100H}. In angle-action coordinates aligned with the stream, the internal components are clearly visible in the subspace $\{\Delta\bar{\theta}_1, \Delta\bar{\varOmega}_1\}$, which corresponds to the angle and frequency aligned with the principal axis of the stream. In contrast, the perpendicular frequencies $\Delta\bar{\varOmega}_2$ and $\Delta\bar{\varOmega}_3$ have a weaker or no internal structure. Their distribution along the stream is shown in Appendix~\ref{App2}.

In the right panel of Figure~\ref{m_loss}, we show the angle and the frequency along the principal axis of the simulated stream. The centre of the cluster is located at the origin, and it is marked as a large grey dot. We also show the area containing 68 per cent of the cluster star distribution with a grey contour line. The stream stars are shown as small black dots for the leading ($\Delta\bar{\theta}_1>0$) and trailing ($\Delta\bar{\theta}_1<0$) arms. These stars are approximately stripped from the centre of the cluster with a frequency in the range $|\Delta\bar{\varOmega}_1|\Approx\Range{0.25}{0.75}$~rad~Gyr$^{-1}$. Once the gravitational potential of the progenitor is negligible, the stripped stars follow a linear motion with constant frequency along the principal axis of the stream (Eq.~5, \citetalias{2025MNRAS.539.2718P}). Since most of the stars are stripped at regular intervals near the pericentres, the stars appear concentrated in peaks. The centres of the three peaks are marked by three coloured dashed lines. Each line marks the frequency of test particles ejected from the cluster at the time of each pericentre passage, with the frequencies distributed uniformly as a function of the angle.

\begin{figure}
\includegraphics[width=1.0\columnwidth]{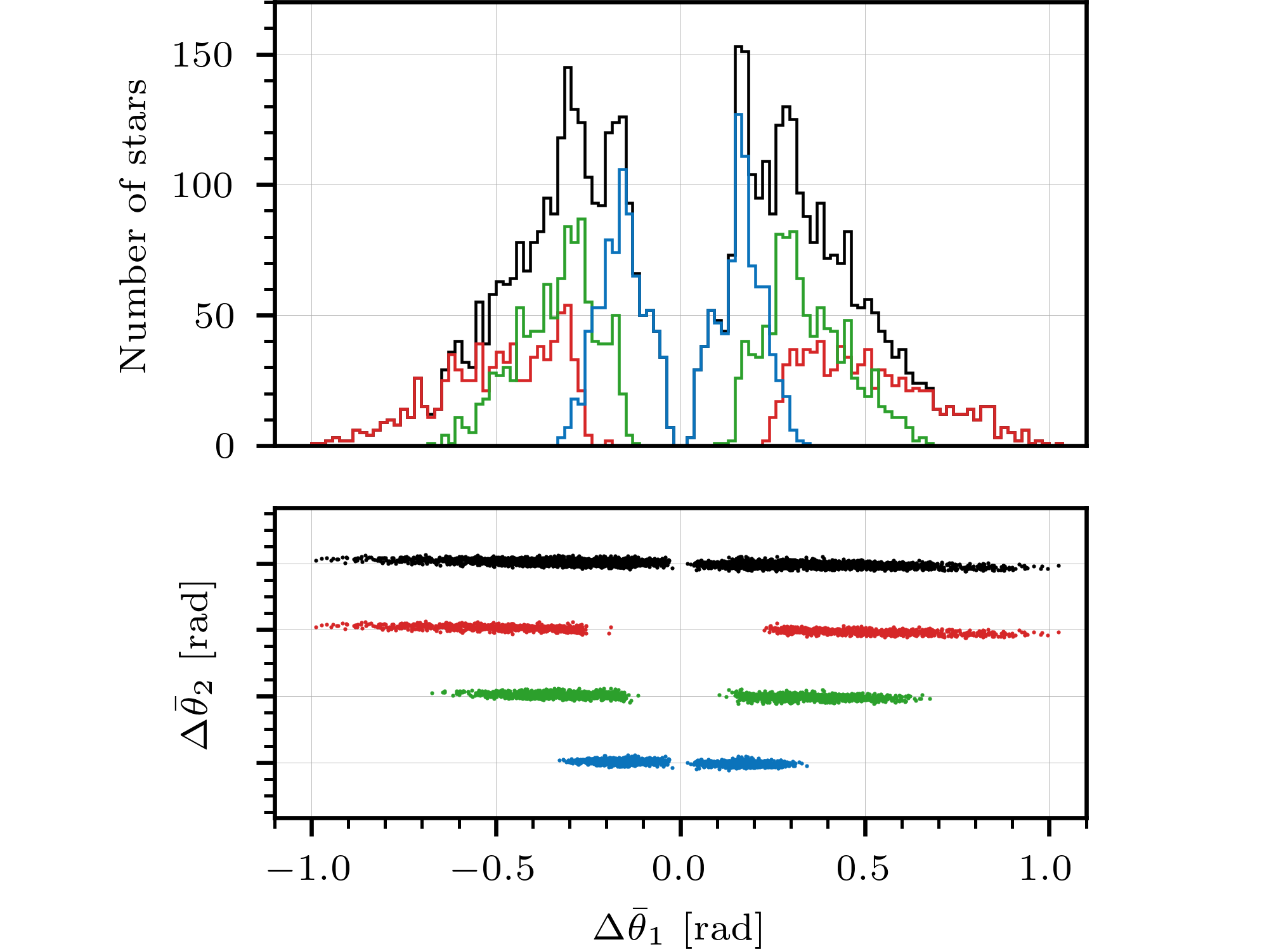}
\caption{\textit{Top:} Number of stream stars as a function of the angle along the principal axis of the stream $\Delta\bar{\theta}_1$ within bins of $18.5$ mrad. The stream stars (black) are divided into three internal components generated by tidal shocks during the first (red), second (green), and third (blue) pericentre passages. \textit{Bottom:} Stream stars and internal components in angle space. The separation between each stream in $\Delta\bar{\theta}_2$ is arbitrary.}
\label{angle_hist}
\end{figure}

Each peak can be separated from the others by grouping the stripped stars into three intervals defined by the time boundaries $[-1500,\,-1055,\,-610,\,-200]$ Myr (App.~D, \citetalias{2025MNRAS.539.2718P}). Each group forms an internal stream. In the top panel of Figure~\ref{angle_hist}, we show the number of stars in each internal component along the principal axis of the stream. Close to the cluster, between $|\Delta\bar{\theta}_1|\Approx\Range{0.1}{0.4}$ deg, we can observe two peaks in the total star distribution (black histogram), corresponding to the third (blue) and second (green) internal streams. The internal streams expand in time in angle space as the stars move along the principal axis. As a result, the density of these streams decreases and they become indistinguishable from the others at large angles. This is the case of the internal stream generated during the first pericentre passage (red). In the bottom panel of Figure~\ref{angle_hist} we show a projection in angle space of each component. The vertical separation between the internal streams is arbitrary and has been chosen to show each component separately. In the next section, we present a methodology for modelling the mass loss of a globular cluster that reproduces this internal structure. We then use this model to describe the internal dynamics of the stream.

\section{Model of the stream in angle-frequency space}\label{model}

In this section, we develop a model of a stellar stream in angle-action coordinates. The main advantage in comparison to an \nbody\ simulation is that this method generates a stream model much faster, especially when the integration time is long and the cluster is composed of a large number of stars. In addition, it allow us to generate large star samples. Using these samples, we can estimate the density along the stream with negligible statistical error and convolve it with observational uncertainties and selection functions. Large samples are in general impractical to obtain with \nbody\ simulations, since they require repeating the simulation many times with different initial conditions. The main drawback of modelling in angle-action coordinates is that the model depends on several free parameters that have to be estimated from a theory or calibrated from a simulation. Furthermore, computing angle-action coordinates requires static potentials with a high degree of symmetry (\S3.5, \citetalias{2008gady.book.....B}).

We model the stream in the reference frame aligned with the principal axes of the stream with the globular cluster centre located at the origin ($\Delta\bar{\theta}, \Delta\bar{\varOmega}$). In these coordinates, the orbit of a stripped star is derived from Eq.~1 of \citetalias{2025MNRAS.539.2718P}:
\begin{equation}\label{int_orb}
\Delta\bar{\theta}_j\var{t} = \kappa\Bigl[ \Delta\bar{\varOmega}_j \;\! (t-t_{\rm s}) + \Delta\bar{\alpha}_j \Bigr] \hfill t\geqslant t_{\rm s},
\end{equation}
where $j=\{1,2,3\}$, $t$ is the simulation time, $t_{\rm s}$ is the stripping time of the stream star (Eq. 8, \citetalias{2025MNRAS.539.2718P}), and the stripping angle is defined as $\Delta\bar{\alpha} \equiv \Delta\bar{\theta}\var{t_{\rm s}}$. In this model, we assume that the leading and trailing arms are symmetric and introduce the parameter $\kappa=(1, -1)$, which gives the sign of the integration for the leading and trailing arms respectively. The orbit of a stream star, as described by Eq.~\ref{int_orb}, is equivalent to a rectilinear uniform motion from an initial position $\Delta\bar{\alpha}$ at a velocity $\Delta\bar{\varOmega}$. For this reason, it is convenient to model the stream in angle-frequency space.

In the following subsections, we model the distribution of stripped stars along the cluster orbit in order to estimate the stripping time of each star. We take the cluster orbit from an initial time $T$ to the present. This initial time $T$ corresponds to the age of the stream and is considered equivalent to the accretion time of the globular cluster. We also model the distribution of stripping angles, which are the initial conditions for the motion, as well as the distribution of frequencies of the stripped stars along the principal axes of the stream. To increase the sample size of the \nbody\ simulation, and therefore to minimise the statistical error when determining the probability distributions, we multiply the angles and frequencies of the stream stars by the parameter $\kappa$. This reverses the sign of the angles and frequencies in the trailing arm, merging the two arms together.

The model of the stellar stream is generated using the following procedure:
\begin{enumerate}
    \setlength\itemsep{0.5em}
    \item Determine the distribution of the number of stripped stars along the cluster orbit from an initial time $t=-T$ to the present time $t=0$.
    \item From this distribution and the total number of stripped stars, generate a random sample of stripping times $t_{\rm s}$.
    \item For each star, randomly generate the initial stripping angle $\Delta\bar{\alpha}$ and frequency $\Delta\bar{\varOmega}$ according to the estimated distributions.
    \item Integrate each star forward in time using Eq.~\ref{int_orb}.
\end{enumerate}

The actions of the stream stars are estimated from their frequencies, which are then used to compute the inverse transformation from angle-actions to Galactocentric cartesian coordinates. This transformation is necessary to evaluate the \textit{Gaia} selection function and simulate observational errors. We then validate the generated model by comparing it with the \nbody\ simulation. Finally, we use the model to study the surface density of the stream as a function of $T$.

\subsection{Distribution of the number of stripped stars along the cluster orbit}\label{num_stripp_stars}

\begin{figure}
\includegraphics[width=1.0\columnwidth]{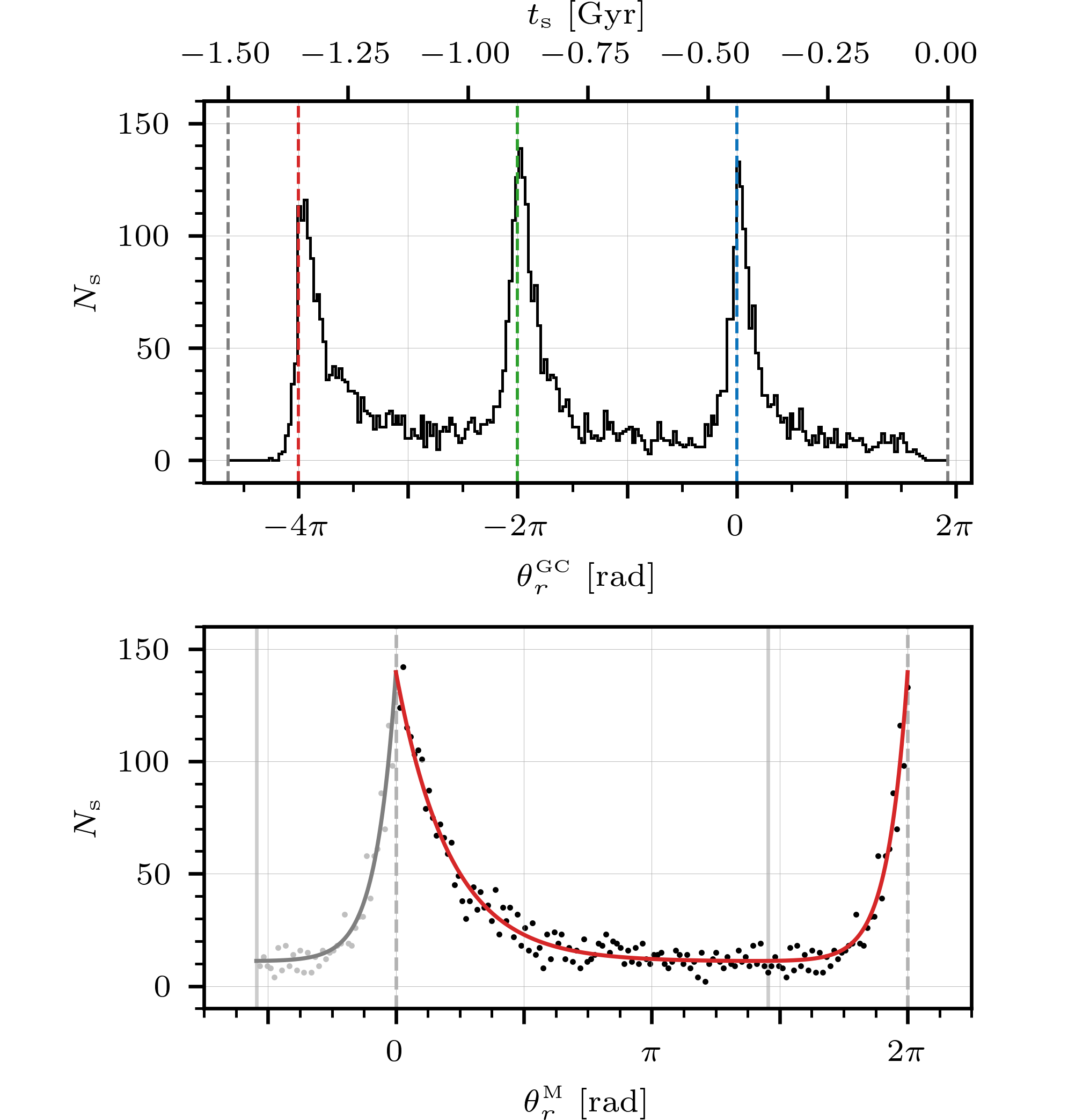}
\caption{\textit{Top:} Histogram of the number of stripped stars $N_{\rm s}$ as a function of the radial angle of the globular cluster $\theta_r^{\GC}$ in bins of $4\pi/139\simeq0.09$~rad. The stripping time $t_{\rm s}$ of the stream stars is indicated on the top coordinate axis. The coloured dashed lines mark the first (red), second (green), and third (blue) pericentre passages, while the grey dashed lines mark the start and end of the simulation. \textit{Bottom:} $N_{\rm s}$ as a function of the radial angle of the globular cluster, corrected for the delay $\theta_r^{\M}$. The angles correspond to the centre of the bins. The red solid line shows the best-fitting double exponential model (Eq.~\ref{double_exp}). The vertical dashed grey lines indicate the limits of the radial period, and the solid grey line indicates the angle at which the two exponentials are equal $\theta_r^{\Lim}$. The second exponential is also shown in grey at the beginning of the period.}
\label{stars_stripped}
\end{figure}

In general, the number of stripped stars depends on the tidal forces acting on the cluster. These forces, in turn, depend on the position of the cluster along its orbit. In angle-action coordinates, the orbit of the cluster is given by Eq.~1 of \citetalias{2025MNRAS.539.2718P}:
\begin{equation}\label{orbit_gc}
\theta_i^{\GC} = \varOmega_i^{\GC}t + \theta_i^{\GC}\var{0},
\end{equation}
where the values of the frequency $\varOmega_i^{\GC}$ and angle at the current time $\theta_i^{\GC}\var{0}$ of the M68 cluster are listed in Eq.~3 of \citetalias{2025MNRAS.539.2718P}. Figure~\ref{stars_stripped} shows the number of stripped stars $N_{\rm s}$ as a function of the radial angle of the globular cluster $\theta_r^{\GC}$, within bins of $4\pi/139\simeq0.09$~rad. The coordinate $\theta_r$ is defined such that the pericentres are located at $\theta_r=2 \pi n$~rad, where $n$ is an integer. Similarly, the apocentres are located at $\theta_r=(2n+1)\pi$~rad. In this simulation, the pericentres correspond to $\theta_r^{\GC}=\{-4\pi, -2\pi,\,0\}$~rad, and the apocentres to $\theta_r^{\GC}=\{-3\pi, -\pi,\,\pi\}$~rad. In Figure~\ref{stars_stripped}, the pericentres are marked with vertical red, green and blue dashed lines. The stripping time $t_{\rm s}$ of the stream stars is also indicated on the upper ordinate axis, and the start and end of the simulation are marked with vertical grey dashed lines.

There is a small delay between the pericentres and the position of the peaks of an angle $\theta_r^{\D}\simeq82.5$~mrad, or equivalently $t_{\D}\simeq6$~Myr. This delay may be because the stars require a period of time to leave the cluster after the tidal shock, or because the stripping time estimation does not account for the influence of the cluster potential on the stripped stars during the stripping process. We introduce the angle $\theta_r^{\M} \equiv \theta_r^{\GC}+\theta_r^{\D}$ to correct the radial angle of the cluster for the delay. When expressed as a function of $\theta_r^{\M}$, the peaks coincide with the beginning and end of the angular radial period. The \nbody\ simulation completed two periods corresponding to $\theta_r^{\M} \in \RangeNI{-4\pi}{0}$. Due to the periodicity of this evolution, the stars stripped within these two periods can be grouped, and $N_{\rm s}$ can be expressed as a single period of $\theta_r^{\M} \in \RangeNI{0}{2\pi}$. In the bottom panel of Figure~\ref{stars_stripped}, we plot the $N_{\rm s}$ as black dots as a function of $\theta_r^{\M}$, where the angle corresponds to the centre of the bin.

We model the evolution of $N_{\rm s}$ using a double exponential function:
\begin{equation}\label{double_exp}
f\var{\theta_r^{\M}} = A\left[\exp\!\left( -\frac{\theta_r^{\M}}{\tau_1}\right ) + \exp\!\left( \frac{\theta_r^{\M}-2\pi}{\tau_2} \right) \right] + C,\end{equation}
where $\theta_r^{\M}\in\RangeNI{0}{2\pi}$. The first exponential approximates the peak after the tidal shock at the pericentre. Its scale length in radians is modelled by the parameter $\tau_1$. The second exponential models the approach to the pericentre, and its scale length is determined by $\tau_2$. The parameter $A$ determines the relative amplitude of the peak with respect to the baseline $C$. We plot the best-fitting $f\var{\theta_r^{\M}}$ as a solid red line in Figure~\ref{stars_stripped}, where the parameters are estimated following the procedure described in Appendix~\ref{App3_par_est}. The values of the parameters are listed in Table~\ref{par_table_a}. In the bottom panel of Figure~\ref{stars_stripped}, we indicate the limits of the radial period with vertical dashed grey lines and mark with a vertical grey solid line the angle where the two exponentials are equal:
\begin{equation}\label{equal_exp}
\theta_r^{\Lim} \equiv \dfrac{2\pi\tau_1}{\tau_1 + \tau_2}.
\end{equation}
We also show in grey the second exponential at the beginning of the period so the asymmetry of the slopes is clearly visible.

\begin{table}
\caption{Parameters of the M68 stellar stream model.}
\begin{subtable}[b]{1.0\columnwidth}

\begin{center}
\vspace{1.25em}
\begin{tabular}{ccccc}
\toprule
&\s{6}$A$&\s{6}$C$&\s{3}$\tau_1$&\s{2}$\tau_2$\\
\midrule
&\s{6}\units{$\ast$}&\s{6}\units{$\ast$}&\s{3}\units{rad}&\s{2}\units{rad}\\[0.27em]
$N_{\rm s}$&\s{6}$128.736$&\s{6}$10.957$&\s{3}$0.660$&\s{2}$0.248$\\[0.27em]
\midrule
&\s{6}\units{mrad Myr$^{-1}$}&\s{6}\units{mrad Myr$^{-1}$}&\s{3}\units{rad}&\s{2}\units{rad}\\[0.27em]
$\mean(\Delta\bar{\varOmega}_1)$&\s{6}$0.295$&\s{6}$0.261$&\s{3}$0.812$&\s{2}$0.456$\\[0.27em]
$\std(\Delta\bar{\varOmega}_1)$&\s{6}$0.099$&\s{6}$0.021$&\s{3}$0.524$&\s{2}$0.451$\\
\bottomrule
\end{tabular}
\end{center}

\caption{Best-fitting parameters of the double exponential model (Eq.~\ref{double_exp}) for the number of stripped stars $N_{\rm s}$, where the units of $A$ and $C$ are: $\ast \equiv {\rm stars}\times(4\pi/139\,{\rm rad})^{-1}$, and mean and standard deviation of the frequency along the principal axis of the stream of the stream stars $\Delta\bar{\varOmega}_1$.}
\label{par_table_a}

\end{subtable}
\begin{subtable}[b]{1.0\columnwidth}

\begin{center}
\vspace{1.25em}
\begin{tabular}{ccc}
\toprule
&$\mu$ \units{mrad}&$\varSigma$ \units{mrad$^2$}\\
\midrule
$\Delta\bar{\alpha}$&
$\begin{pmatrix}
-0.103 \\
-1.602 \\
-2.255 \\
\end{pmatrix}$&
$\begin{pmatrix}
\phantom{0}0.088 \!&\! \phantom{0}0.618 \!&\! \phantom{0}0.684\\
\phantom{0}0.618 \!&\! 27.749 \!&\! 13.583\\
\phantom{0}0.684 \!&\! 13.583 \!&\! 25.398\\
\end{pmatrix}$\\
\bottomrule
\end{tabular}
\end{center}

\caption{Mean and covariance matrix of the best-fitting multivariate Gaussian distribution to the stripping points in angle space.}
\label{par_table_b}

\end{subtable}
\begin{subtable}[b]{1.0\columnwidth}

\begin{center}
\vspace{1.25em}
\begin{tabular}{ccc}
\toprule
&$\mu$&$\sigma$\\
&\units{mrad Gyr$^{-1}$}&\units{mrad Gyr$^{-1}$}\\
\midrule
$\Delta\bar{\varOmega}_2$&$-2.465$&$4.150$\\[0.27em]
$\Delta\bar{\varOmega}_3$&$-6.883$&$4.239$\\
\bottomrule
\end{tabular}
\end{center}

\caption{Mean and standard deviation of the best-fitting univariate Gaussian distribution to the frequencies perpendicular to the principal axis of the stream.}
\label{par_table_c}

\end{subtable}
\label{par_table}
\end{table}

The probability of a star being stripped as a function of $\theta_r^{\M}$ is obtained by normalising Eq.~\ref{double_exp}:
\begin{equation}\label{stripped_stars_pdf}
p\var{\theta_r^{\M} \:\!|\:\! A, C, \tau_1, \tau_2} = N_{\rm c} \, f\var{\theta_r^{\M}},
\end{equation}
where $N_{\rm c}$ is the normalisation constant. This probability density function (PDF) is periodic, and can therefore be generalised for an orbit of arbitrary length by wrapping the angle $\theta_r^{\M}$ within the interval $\RangeNI{0}{2\pi}$. In Appendix~\ref{App3_norm_const}, we provide a formal definition of the normalisation constant for a general orbit. Using Eq.~\ref{stripped_stars_pdf}, we can generate random samples of stripping angles of the stream stars $\theta_r^{\St}$. In Appendix~\ref{App3_rdm_sample}, we introduce the methodology for generating random samples from this distribution. The stripping time of the stream stars $t_{\rm s}$ is obtained by subtracting the delay $\theta_r^{\D}$ from $\theta_r^{\St}$ and isolating the time from Eq.~\ref{orbit_gc}:
\begin{equation}
t_{\rm s} = \dfrac{ \theta_r^{\St}-\theta_r^{\GC}\var{0} }{ \varOmega_r^{\GC} } - t_{\D}.
\end{equation}

Given an orbit of duration $T$, we define the number of stripped stars per unit of time per arm as $N_{T}$. The total number of stripped stars along the orbit is therefore given by:
\begin{equation}
N_{\St} = 2 \!\; T \;\!\! N_{T}.
\end{equation}
For the \nbody\ simulation of M68, the time between the two complete periods is approximately $913.95$ Myr, and the total number of stripped stars is $4071$. Assuming these stars are equally distributed between the two arms, we fix the parameter $N_{T} \simeq 2.23$~stars~Myr$^{-1}$~arm$^{-1}$ to reproduce the results of the simulation. In Section~\ref{estimations}, we determine the value of this parameter from the observational data of M68.

The total mass loss is estimated by multiplying $N_{\St}$ by the mean of the star mass distribution of the cluster. The synthetic population of the cluster follows a Kroupa initial mass function with a mean of approximately $0.289$~M$_{\Sun}$ (\S2.1, \citetalias{2025MNRAS.539.2718P}). For the stream simulated with this model, the estimated mass lost is about $1930.94$~M$_{\Sun}$ for $T=1.5$~Gyr, which corresponds to $\Approx1.55$ per cent of the initial mass of the cluster (Table~1, \citetalias{2025MNRAS.539.2718P}). In this model, we assume that this mass loss is small and neglect variations in mass and density distribution of the cluster along the orbit. We also neglect the variations in the stellar mass function along the stream caused by star mass segregation \citep{2022MNRAS.510..774W}. We assume that this simplification is valid for orbits of a similar duration to that of the simulation.

\subsection{Stripping points distribution}\label{strip_dist}

We determine the distribution of stripping points $\Delta\bar{\alpha}$ by integrating the orbit of the stream stars backwards in time until they return to the globular cluster. The integration time is estimated using the ratio of the relative angle and frequency of each stream star to the globular cluster (Eq.~7 of \citetalias{2025MNRAS.539.2718P}). This procedure is analogous to the Inverse Time Integration (\texttt{invi}) method introduced in \S4 of \citetalias{2025MNRAS.539.2718P}. The resulting distribution is roughly flat along the principal axis of the stream and approximately a correlated Gaussian in the perpendicular plane (Fig.~4, \citetalias{2025MNRAS.539.2718P}).

\begin{figure}
\includegraphics[width=1.0\columnwidth]{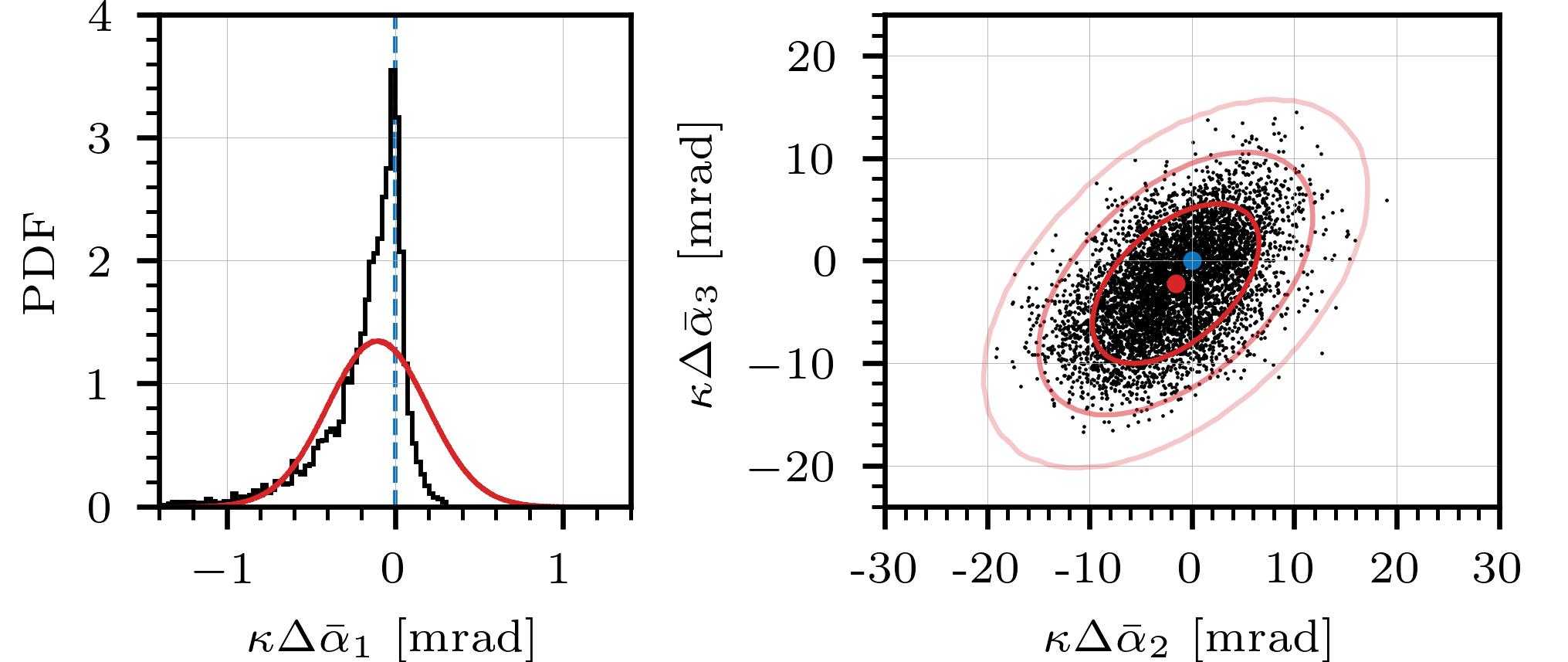}
\caption{\textit{Left:} Distribution of stripping points along the principal axis of the stream $\Delta\bar{\alpha}_1$ (black) and the first component of the best-fitting multivariate Gaussian distribution (red). The symbol $\kappa$ indicates that both arms are shown together. The vertical dashed blue line marks the position of the cluster centre. \textit{Right:} Small black dots show the positions of the stripping points in the plane perpendicular to the stream ($\Delta\bar{\alpha}_2, \Delta\bar{\alpha}_3$). The red lines show the $1$, $2$, and $3\text{-}\sigma$ levels of the best-fitting multivariate Gaussian distribution. The large red dot indicates the median of the distribution and the large blue dot marks the cluster centre.}
\label{invi_alpha}
\end{figure}

We model this distribution using a multivariate Gaussian distribution. Due to the presence of outliers in the coordinate $\Delta\bar{\alpha}_1$ (Left panel Fig.~4, \citetalias{2025MNRAS.539.2718P}), we estimate the mean using the median. The values of the parameters of the best-fitting distribution are listed in Table~\ref{par_table_b}. In the left panel of Figure~\ref{invi_alpha}, we plot in black the distribution of stripping points along the principal axis of the stream $\Delta\bar{\alpha}_1$, estimated with an histogram. We also plot the first component of the best-fitting multivariate Gaussian distribution as a solid red line. Due to the asymmetric distribution of $\Delta\bar{\alpha}_1$, the Gaussian model overestimates the positive values and underestimates the peak. However, these differences are negligible because variations in $\Delta\bar{\alpha}\Approx0.5$~mrad are much smaller than the length of an arm of a stream. In the case of the \nbody\ simulation of M68, the length of an arm is $\Approx1$~rad (Fig~3c, \citetalias{2025MNRAS.539.2718P}). Nevertheless, in order to provide a complete theoretical study, we present a list of distributions that better fit $\Delta\bar{\alpha}_1$ in Appendix~\ref{App4}.

In the right panel of Figure \ref{invi_alpha}, we plot the stripping points in the perpendicular plane of the stream ($\Delta\bar{\alpha}_2, \Delta\bar{\alpha}_3$). The $1$, $2$, and $3$-$\sigma$ levels of the best-fitting multivariate Gaussian distribution are shown as solid red lines. The median of the distribution is marked with a large red dot and the centre of the globular cluster is marked with a large grey dot. As can be seen, this model provides a good fit to the distribution of stripping points in this plane.

\subsection{Frequency distribution along the stream}\label{freq_1_dist}

In the reference frame aligned with the principal axes of the stream, the frequencies are approximately proportional to the actions, according to the following relation:
\begin{equation}\label{linear_relation}
\Delta\bar{\varOmega}_i \approx \lambda_i \:\! \Delta\bar{J}_i,
\end{equation}
where $\lambda_i$ are the eigenvalues of a linear approximation of the Hamiltonian (Eq.~4, \citetalias{2025MNRAS.539.2718P}). Note that this is only valid for a progenitor of size and mass of a globular cluster \citep{2015MNRAS.450..575A}. For the globular cluster M68 and the potential model of the Milky Way used in the \nbody\ simulation, the absolute values of the eigenvalues satisfy $\lambda_1 \gg \lambda_2, \lambda_3$ (Eq.~6, \citetalias{2025MNRAS.539.2718P}). Since the three actions of the stream stars are similar in magnitude (Fig. 3a, \citetalias{2025MNRAS.539.2718P}), it follows that $\Delta\bar{\varOmega}_1 \gg \Delta\bar{\varOmega}_2, \Delta\bar{\varOmega}_3$. Consequently, the length and the surface density of the stream are determined by $\Delta\bar{\varOmega}_1$. We therefore model this frequency in more detail, on the assumption that it is uncorrelated with the frequencies perpendicular to the stream.

\begin{figure}
\includegraphics[width=1.0\columnwidth]{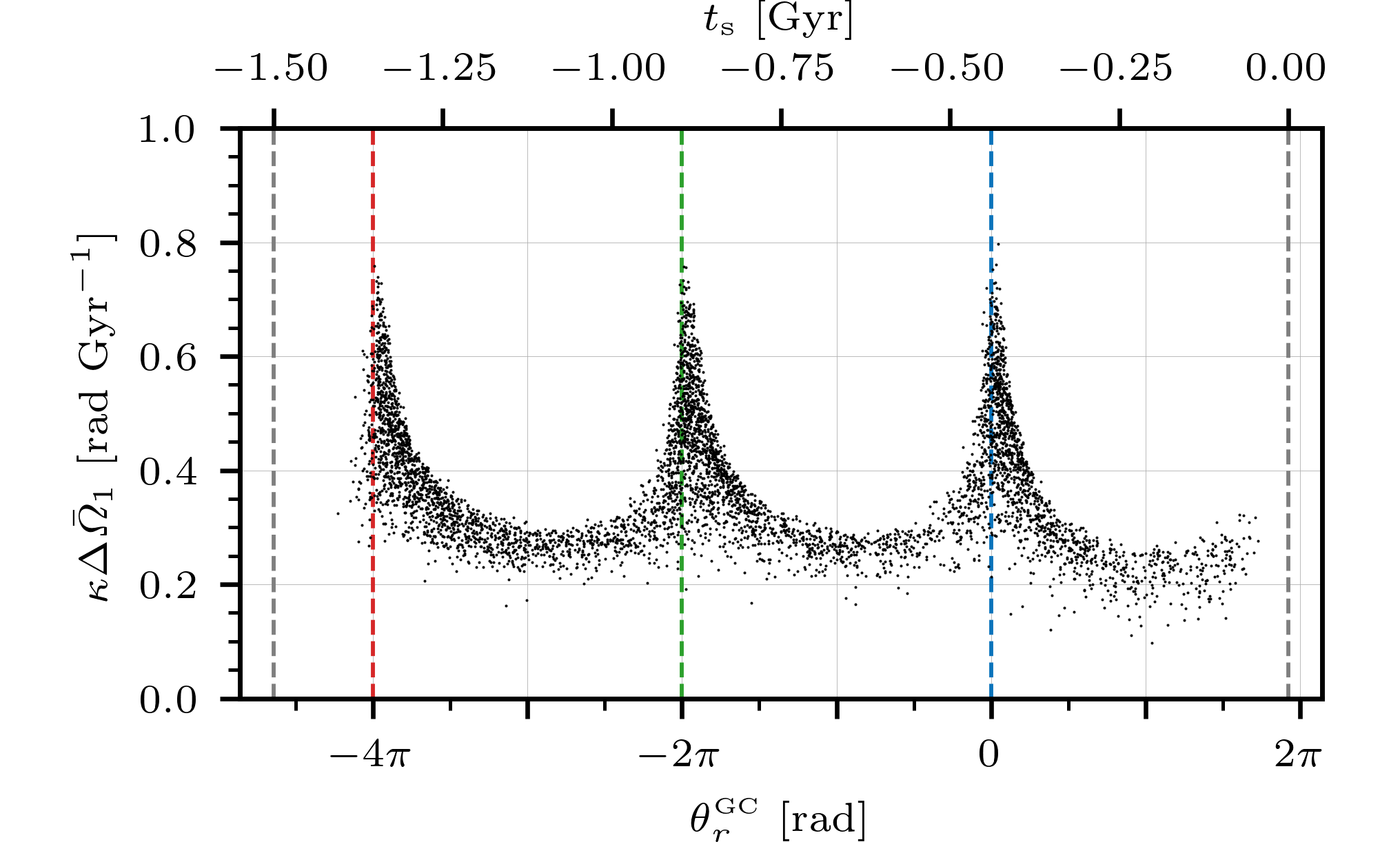}
\caption{Frequency along the principal axis of the stream of the stripped stars $\Delta\bar{\varOmega}_1$ as a function of the radial angle of the globular cluster $\theta_r^{\GC}$. The symbol $\kappa$ indicates that both arms are shown together. The stripping time $t_{\rm s}$ of the stream stars is indicated on the top coordinate axis. The coloured dashed lines mark the first (red), second (green), and third (blue) pericentre passages, while the grey dashed lines mark the start and end of the simulation.}
\label{Ar_F1}
\end{figure}

In Figure~\ref{Ar_F1}, we plot the dependency of the frequency along the principal axis of the stream of the stripped stars $\Delta\bar{\varOmega}_1$ on the radial angle of the globular cluster $\theta_r^{\GC}$. We observe that, on average, the frequencies of the stars stripped at pericentres are larger than those stripped at apocentres. Similarly, the dispersion of the distribution also changes along the cluster orbit, being maximum at the pericentres and minimum at the apocentres. This is because the tidal forces are stronger at the pericentres, resulting in stars being stripped with greater relative velocity with respect to the cluster. The frequency peaks appear with a delay relative to the pericentres, which are indicated by vertical coloured dashed lines. We therefore express the frequency as a function of the corrected angle $\theta_r^{\M}$, similarly as we do for the number of stripped stars (Section~\ref{num_stripp_stars}).

\begin{figure*}
\includegraphics[width=1.0\textwidth]{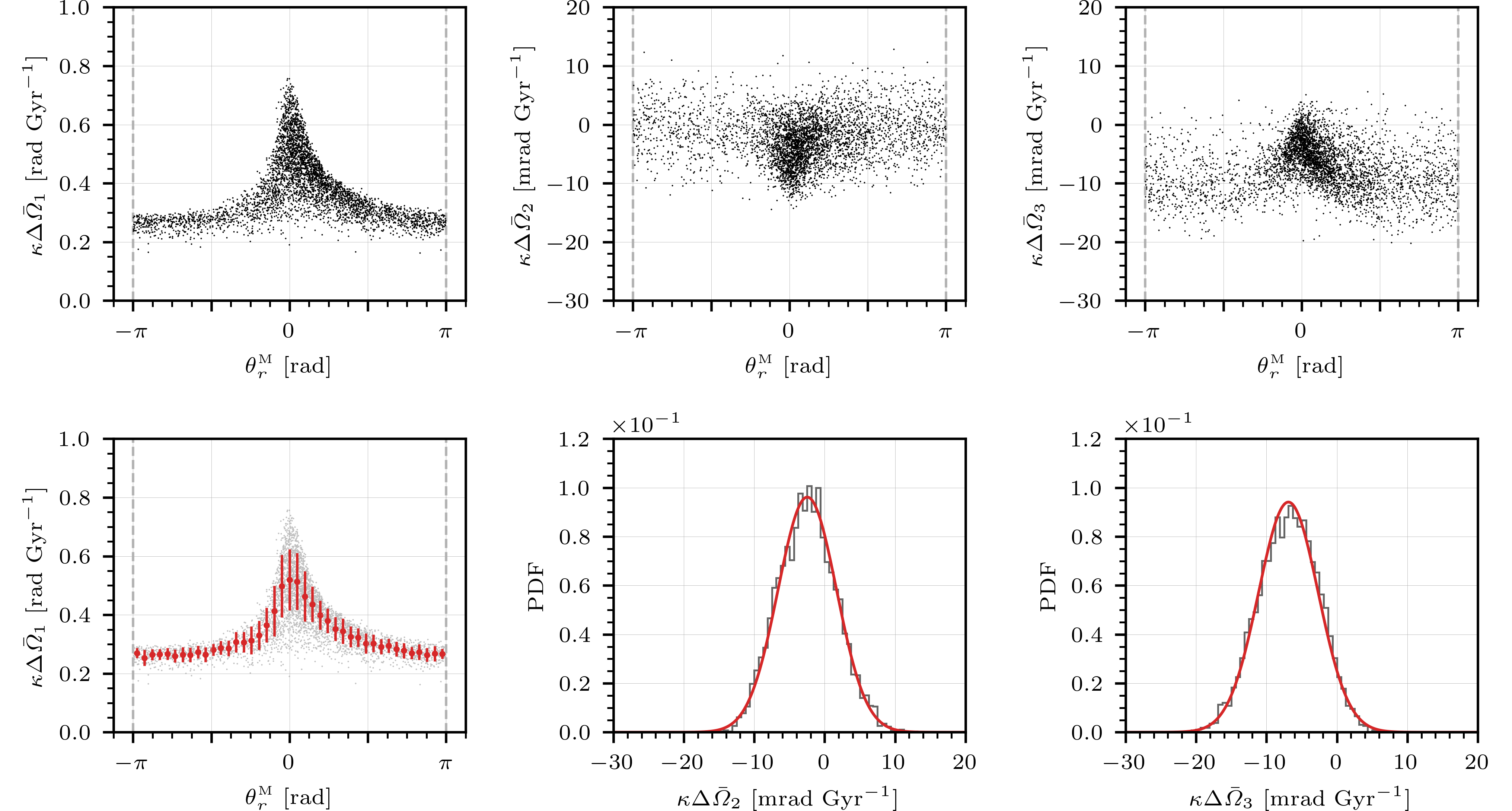}
\caption{\textit{Top:} Frequency of the stripped stars along the principal axis of the stream $\Delta\bar{\varOmega}_1$ (\textit{Left}), $\Delta\bar{\varOmega}_2$ (\textit{Centre}), and $\Delta\bar{\varOmega}_3$ (\textit{Right}) as a function of the radial angle of the globular cluster corrected for the delay $\theta_r^{\M}$. The vertical dashed grey lines indicate the boundaries of the period. \textit{Bottom left:} Same as the top left panel, but with grey dots. The large red dots with error-bars indicate the mean and the standard deviation of the stripped stars within bins of $\theta_r^{\M}\simeq0.153$~rad, where the angle is taken as the centre of the bin. \textit{Bottom centre:} Distribution of the perpendicular frequency to the stellar stream of the stripped stars $\Delta\bar{\varOmega}_2$, estimated using a histogram (grey line). The best-fitting Gaussian distribution is shown in red. \textit{Bottom right:} Same as bottom centre, but for $\Delta\bar{\varOmega}_3$. In all panels, the symbol $\kappa$ indicates that both arms are shown together.}
\label{peaks_distributions}
\end{figure*}

In the top left panel of Figure~\ref{peaks_distributions}, we plot the frequency $\Delta\bar{\varOmega}_1$ of the stars stripped during the angular interval $\theta_r^{\M}\in\Range{-4\pi}{0}$. This interval corresponds to the two complete periods present in the simulation. The angle is wrapped within $\theta_r^{\M} \in \RangeNI{-\pi}{\pi}$, with the pericentre at $\theta_r^{\GC}=0$~rad. To visualise the variations in the mean and standard deviation (std) of the frequency along the cluster orbit, we divide $\theta_r^{\M}$ in 41 bins of approximately $0.153$~rad and compute the mean and standard deviation of $\Delta\bar{\varOmega}_1$ of the stars within each bin. We plot the mean and standard deviation as a large red dot with error-bars in the bottom left panel of Figure~\ref{peaks_distributions}. We also show the frequency values as small grey dots for reference. We observe that the mean and the standard deviation are approximately constant close to the apocentre at $\theta_r^{\M}=\{-\pi,\,\pi\}$~rad, and increase at the pericentre. We also note that the slopes of the peak are asymmetric, with the incident slope being steeper than the exiting.

\begin{figure}
\includegraphics[width=1.0\columnwidth]{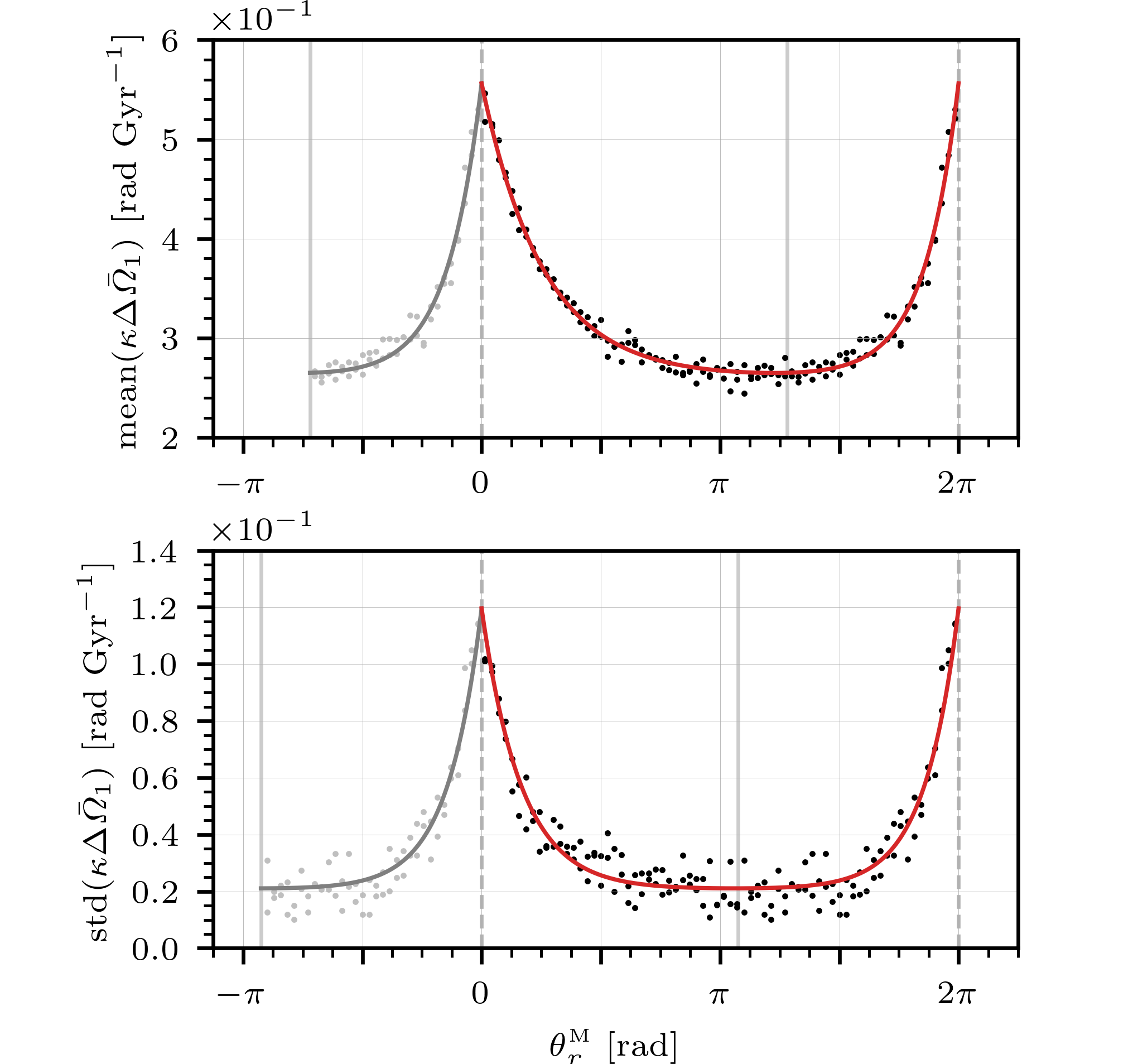}
\caption{\textit{Top:} Mean of the frequency along the principal axis of the stream $\Delta\bar{\varOmega}_1$ as a function of the radial angle of the globular cluster corrected for the delay $\theta_r^{\M}$. The means are calculated for the stars within bins of $2\pi/70 \simeq 0.09$~rad in $\theta_r^{\M}$, with the angles corresponding to the centre of the bins. The symbol $\kappa$ indicates that both arms are shown together. The red solid line shows the best-fitting double exponential model (Eq.~\ref{double_exp}). The vertical dashed grey lines indicate the limits of the radial period, and the solid grey line indicates the angle at which the two exponentials are equal $\theta_r^{\Lim}$ (Eq.~\ref{equal_exp}). The second exponential is also shown in grey at the beginning of the period. \textit{Bottom:} Same as top panel, but showing the standard deviation of $\Delta\bar{\varOmega}_1$.}
\label{mean_std_exp_model}
\end{figure}

In order to model the evolution of the distribution of $\Delta\bar{\varOmega}_1$, we divide the interval $\theta_r^{\M} \in \RangeNI{-\pi}{\pi}$ into 70 bins of $2\pi/70 \simeq 0.09$~rad, and compute the mean and standard deviation of the frequency for the stars within each bin. In the top and bottom panels of Figure~\ref{mean_std_exp_model}, we plot the results for the mean and standard deviation respectively, where the angle $\theta_r^{\M}$ is taken as the centre of the bin. In the case of the mean, the data shows a slightly asymmetric peak with low dispersion along the radial angle. On the other hand, the standard deviation peak is approximately symmetric and the dispersion is larger. We model the evolution of the mean and standard deviation using the double exponential function introduced in Eq.~\ref{double_exp}. We calculate the best-fitting parameters using the method described in Appendix~\ref{App3_par_est}, and list the results in Table~\ref{par_table_a}. The best-fitting models are shown as a solid red lines in Figure~\ref{mean_std_exp_model}. We observe that the models are in very good agreement with the data obtained from the \nbody\ simulation.

We model the frequency distribution of $\Delta\bar{\varOmega}_1$ of the stripped stars within each bin as a Gaussian distribution, with the mean and standard deviation given by the best-fitting double exponential models. We verify this assumption in Appendix~\ref{App4} and propose a more accurate model that depends on additional free parameters. Thus, the frequency distribution of $\Delta\bar{\varOmega}_1$ depends on the stripping angle of the stream stars along the cluster orbit. For a given unwrapped stripping angle $\theta_r^{\St}$, computed using the method described in Section~\ref{num_stripp_stars}, we evaluate the double exponential models for the remainder, which is defined as follows:
\begin{equation}
\theta_r^{\RS} \equiv \theta_r^{\St} \;\!\reminder\;\! 2\pi.
\end{equation}

\subsection{Frequency distributions perpendicular to the stream}

In the top centre and top right panels of Figure~\ref{peaks_distributions}, we plot the frequencies of the stripped stars along the axes perpendicular to the stream $\Delta\bar{\varOmega}_2$ and $\Delta\bar{\varOmega}_3$. These plots are analogous to the plot for $\Delta\bar{\varOmega}_1$ in the top left panel of the same figure. The prevalence of negative frequencies is a consequence of the choice of a right-handed basis for a clockwise rotating cluster, or equivalently for a cluster of $J_{\phi} \equiv L_z<0$. The two frequencies present a peak at $\theta_r^{\M}=0$, corresponding to the pericentre. These peaks are less prominent than the peak of $\Delta\bar{\varOmega}_1$. Similarly, they are more populated than the rest of the period due to the greater mass loss near the pericentres (Section~\ref{num_stripp_stars}).

The perpendicular frequencies are about $50$ times smaller than $\Delta\bar{\varOmega}_1$ (Section~\ref{freq_1_dist}), so their impact on the length and surface density of the stream is less significant. Therefore, we neglect the variations in $\theta_r^{\M}$ and model the frequencies integrated within the interval $\theta_r^{\M} \in \RangeNI{-\pi}{\pi}$ with Gaussian distributions. The distributions of $\Delta\bar{\varOmega}_2$ and $\Delta\bar{\varOmega}_3$, estimated with a histogram, are shown as grey lines in the bottom centre and bottom right panels of Figure~\ref{peaks_distributions}. We also plot the best-fitting univariate Gaussian distributions for each frequency in red and list the values of the best-fitting parameters in Table~\ref{par_table_c}. We verify the compatibility of the Gaussian model with the simulated data using a Kolmogorov-Smirnov test. We obtain $\text{p\:\!-values}$ of approximately $0.44$ and $0.12$ for $\Delta\bar{\varOmega}_2$ and $\Delta\bar{\varOmega}_3$, respectively. This indicates that the hypothesis that the data are generated by Gaussian distributions is not rejected at the 95 per cent confidence level ($\alpha = 0.05$).

\subsection{Actions and inverse coordinate transformation}\label{actions_inv_trans}

In the reference frame aligned with the principal axes of the stream, the actions are proportional to the frequencies, as given by Eq.~\ref{linear_relation}. The proportionality constants are the eigenvalues. These can be evaluated at the position of the globular cluster using the Torus Mapper (TM) algorithm\footnote{Galpy \citep{2015ApJS..216...29B} provides a Python front end for this functionality} \citep{2016MNRAS.456.1982B}. Alternatively, the eigenvalues can be estimated from the \nbody\ simulation, as described in Appendix~F of \citetalias{2025MNRAS.539.2718P}. The actions are required for the inverse transformation from angle-action coordinates to Galactocentric cartesian coordinates. This transformation is also computed numerically using the Torus Mapper code. With a tolerance $\texttt{tol}=2\pd{-4}$, the precision achieved in sky coordinates is $\Approx0.2$~deg from the true value.

\begin{figure}
\includegraphics[width=1.0\columnwidth]{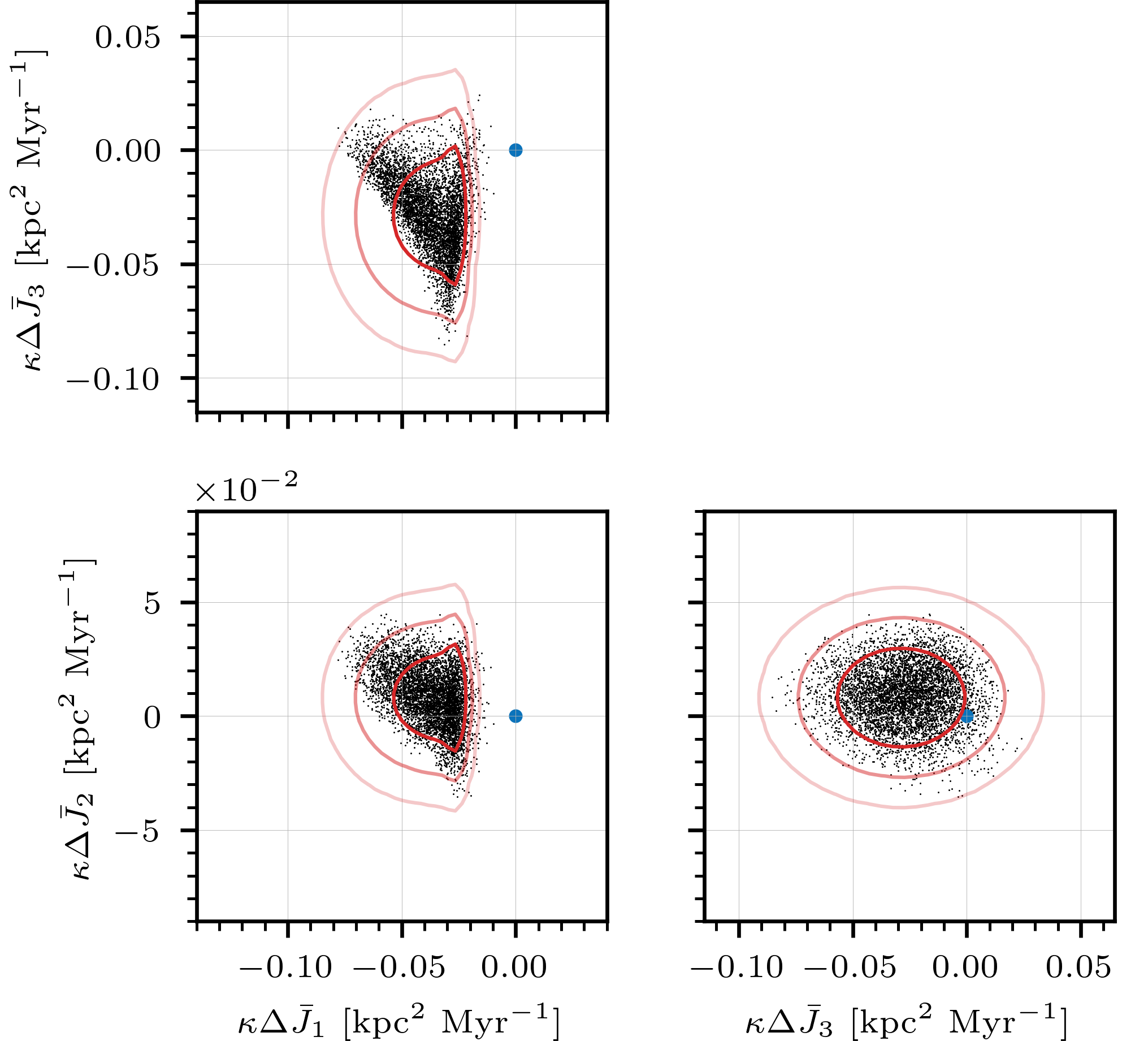}
\caption{Distribution of actions of the stream stars relative to the cluster centre in the reference frame aligned with the principal axes of the stream (small black dots). The large blue dot indicates the cluster centre. The red lines show the $1$, $2$, and $3\text{-}\sigma$ levels of the distributions of actions obtained from the model of the frequencies. The symbol $\kappa$ indicates that both arms are shown together. \textit{Left:} Actions along the principal axis of the stream $\Delta\bar{J}_1$, versus the perpendicular actions $\Delta\bar{J}_2$ and $\Delta\bar{J}_3$. \textit{Right:} Perpendicular actions to the stream.}
\label{actions}
\end{figure}

In Figure~\ref{actions}, we plot the actions of the stream stars of both arms as a small black dots, and the position of the cluster centre as a large blue dot. The $1$, $2$, and $3\text{-}\sigma$ levels of the distribution obtained from the model of the frequencies are indicated as red lines. The left panels show the actions along the principal axis of the stream $\Delta\bar{J}_1$, versus the perpendicular actions $\Delta\bar{J}_2$ and $\Delta\bar{J}_3$. We note that these variables are correlated. These correlations have been neglected when assuming that the frequency distributions are uncorrelated (Section~\ref{freq_1_dist}). Nevertheless, the model accurately describes the central tendency and variance of the simulated stars. On the other hand, the perpendicular frequencies to the stream are uncorrelated, and the model accurately describes their distribution.

\subsection{Comparison with the \texorpdfstring{\textit{N}\!\!\:-body}{N-body} simulation}

\begin{figure}
\includegraphics[width=1.0\columnwidth]{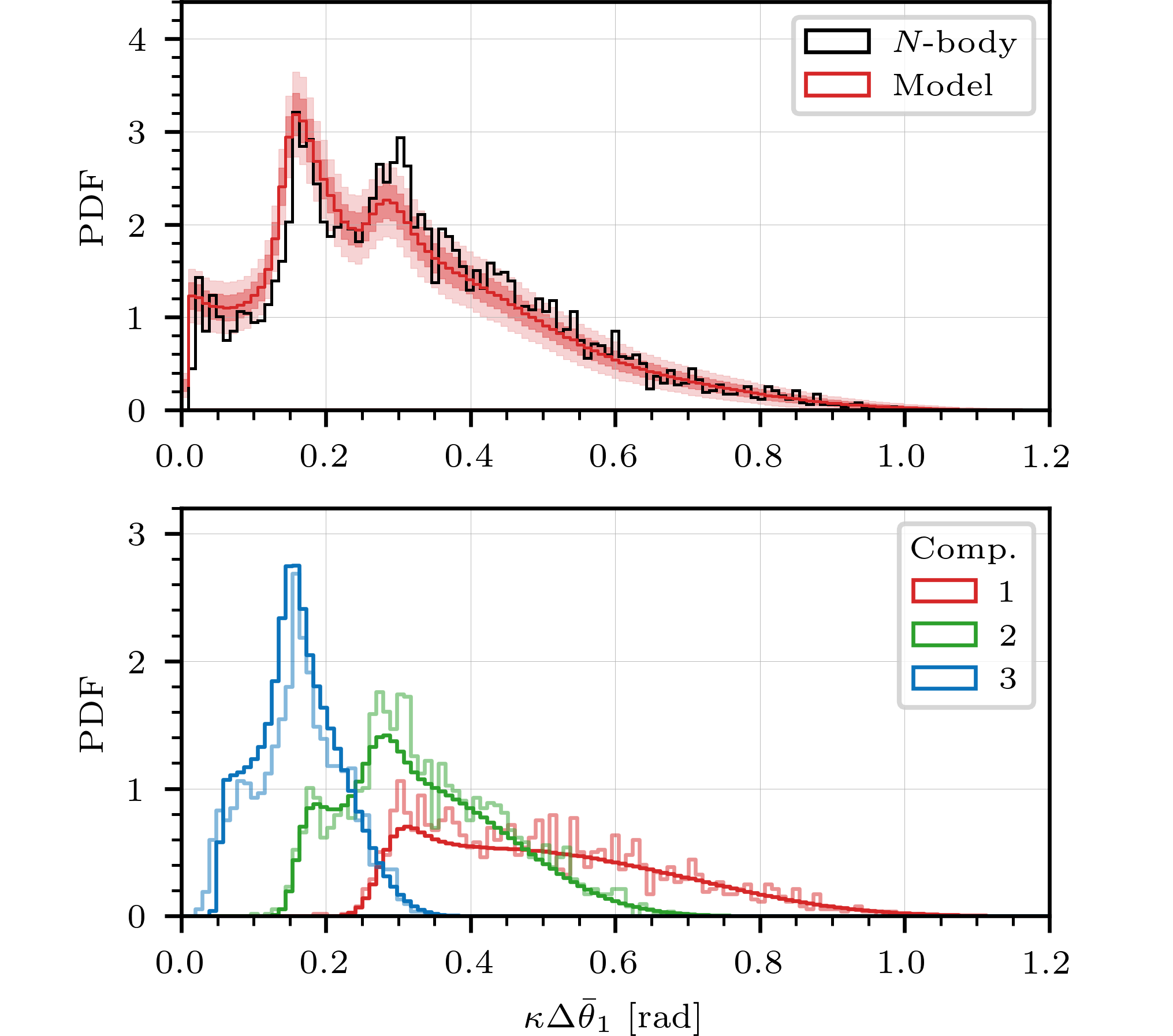}
\caption{\textit{Top:} Mean distribution of angles along the principal axis of the stream $\Delta\bar{\theta}_1$ for $10^3$ models that mock the \nbody\ simulation (red line). The $1$ and $2\text{-}\sigma$ levels are marked with light shaded red areas. The black histogram shows the distribution of stars of the \nbody\ simulation. The distributions are estimated by taking both arms together and using a histogram with $6/625$~rad bins. \textit{Bottom:} Distributions for the internal components of the stream, corresponding to the first (red), second (green), and third (blue) pericentre passages. The coloured lines mark the mean of the models, and the light coloured lines mark the \nbody\ simulation.}
\label{model_comparison}
\end{figure}

We validate the model by comparing it with the \nbody\ simulation of the M68 globular cluster. We simulate $10^3$ streams of $T=1.5$~Gyr and select the stars within $\theta_r^{\GC}\in \Range{-4\pi}{7\pi/4}$~rad or equivalently $t_{\rm s} \in \Range{-1353.28}{-39.62}$~Myr. This approximately corresponds to the range in which the stars are stripped in the \nbody\ simulation (Section~\ref{mass_loss}). We estimate the distribution of angles along the principal axis of the stream $\Delta\bar{\theta}_1$ of each sample using a histogram with bins of $6/625$~rad. In the top panel of Figure~\ref{model_comparison}, we plot the mean of the distributions in red and mark the $1$ and $2\text{-}\sigma$ levels with a shaded red area. We also plot the histogram for the stars of the \nbody\ simulation in black for reference. The bottom panel shows the distributions of the internal components of the stream, as defined in Section~\ref{mass_loss}. Here, we mark the mean of the model with coloured lines and the \nbody\ with light coloured lines. The components follow the same colour code as in Figure~\ref{angle_hist}.

We observe a slight overestimation of the first peak, which corresponds to the last pericentre passage (blue). We note that we have not used data within $\theta_r^{\GC} \in \RangeNI{0}{2\pi}$~rad to calibrate the model, since the last period is incomplete. As shown in Figure~\ref{Ar_F1}, the stars stripped during this period have a slightly lower mean and a slightly larger standard deviation of $\Delta\bar{\varOmega}_1$ than the stripped stars in previous periods. 
We also observe a slight underestimation of the second peak, which corresponds to the second pericentre passage (green) plus the slowest stars from the first pericentre passage (red). The \nbody\ simulation is within the $2\text{-}\sigma$ level of the model, so this discrepancy could be explained by a statistical bias in the \nbody\ simulation sample. Alternatively, a slight overestimation of the standard deviation of $\Delta\bar{\varOmega}_1$ makes the stars expand faster and therefore appear less compact near the peak. We consider these differences to be negligible and conclude that the model is in very good agreement with the simulation.

\subsection{Dynamical evolution}

\begin{figure}
\includegraphics[width=1.0\columnwidth]{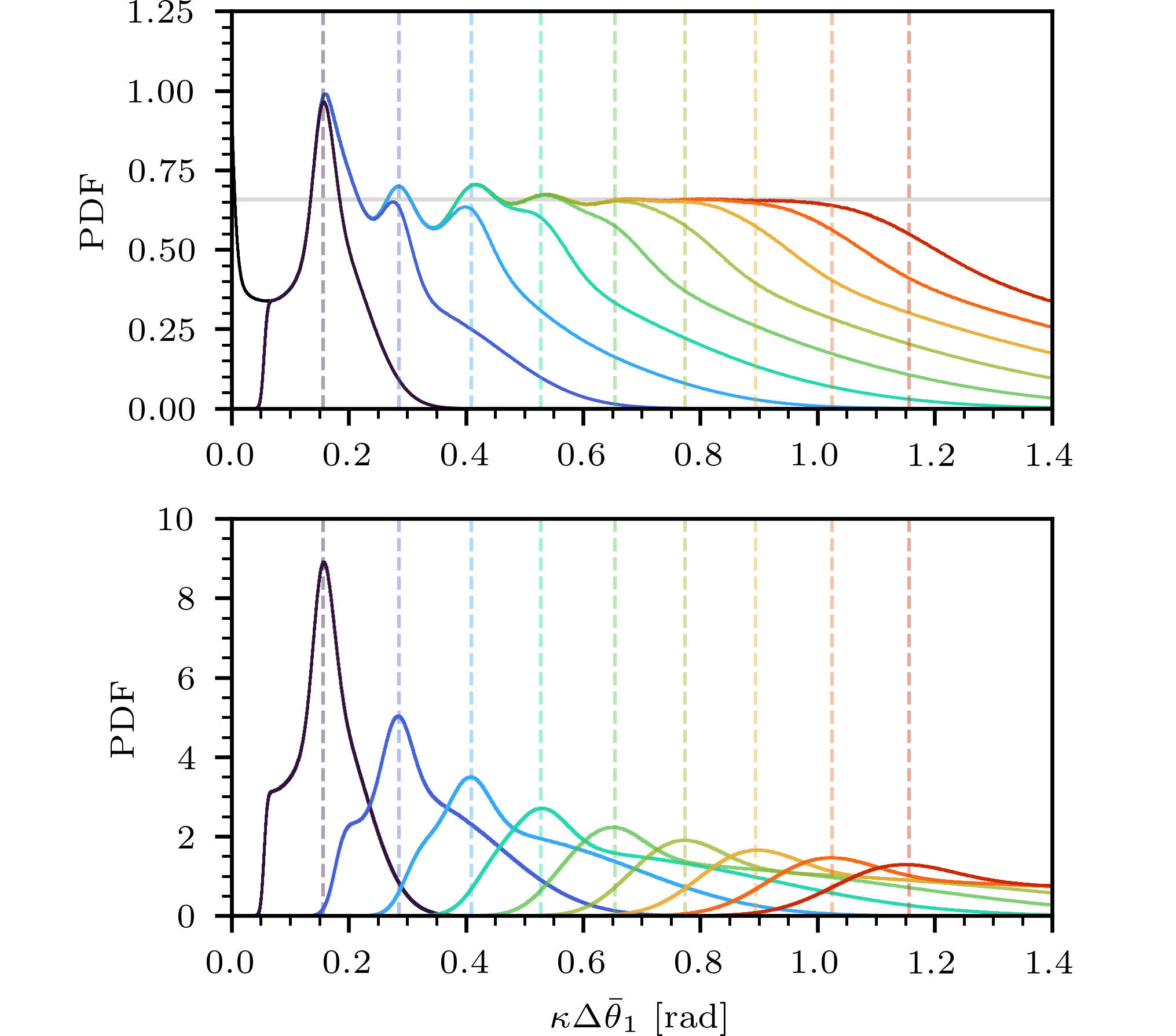}
\caption{Density distribution along the principal axis of a simulated stream $\Delta\bar{\theta}_1$ taking both arms together. The progenitor cluster undergoes nine complete angular periods $\theta_r^{\GC} \in \RangeNI{-17\pi}{0}$. Each period is colour-coded, with red representing the first period and dark purple representing the last. The vertical dashed lines indicate the angle corresponding to the maximum value of the distribution of each period. \textit{Top:} The distributions include the stars stripped up to the period indicated by the colour of the line. The distribution of stars stripped between $\theta_r^{\GC} \in \RangeNI{\pi}{\theta_r^{\Z}}$~rad is shown in black at $\Delta\bar{\theta}_1\Approx\Range{0}{0.15}$~rad. The grey horizontal line marks the density saturation value. \textit{Bottom:} Distributions of each period separately.}
\label{model_evolution}
\end{figure}

This model allow us to describe the evolution of a stream from an arbitrary accretion time $T$. In Figure~\ref{model_evolution}, we show an stream of nine complete angular periods, with limits $\theta_r^{\GC} = -(2n+1)\pi$~rad, where $n$ is an integer such that $-1\leqslant n \leqslant 8$. These periods are indicated by a colour gradient, with red representing the first period and dark purple representing the last. This angular length corresponds to an accretion time $T\simeq4323.29$~Myr. In the top panel, we show the distribution of stripped stars along the principal axis of the stream $\Delta\bar{\theta}_1$ up to the period indicated by the colour of the line. We also include the stars stripped during $\theta_r^{\GC} \in \RangeNI{\pi}{\theta_r^{\Z}}$~rad in black, located at $\Delta\bar{\theta}_1\Approx\Range{0}{0.15}$~rad. In the bottom panel we show each period separately. In both panels, the dashed vertical lines indicate the angle corresponding to the maximum value of the distribution of each period.

In general, the stream stars are stripped in bursts during the tidal shocks at the pericentres (Section~\ref{mass_loss} and \ref{num_stripp_stars}). These stars form peaks in angular space $\Delta\bar{\theta}_1$ that expand along the principal axis of the stream. The stream stars stripped close to the pericentre have large frequencies $\Delta\bar{\varOmega}_1$ with high dispersion (Section~\ref{freq_1_dist}), so they move and expand quickly. On the other hand, the stars stripped closer to the apocentre have lower frequencies with lower dispersion and remain close together for a longer period of time. In the top panel of Figure~\ref{model_evolution}, we can see the peak corresponding to the last pericentre passage (dark purple). In this case, the faster stars have not had time to expand, so these stars together with the slowest form a clear overdensity. The section between the peak and the cluster, which is located at the origin of coordinates, has a density of about one third that of the peak. This region is therefore visible as an underdensity.

As can be seen in the bottom panel, the stars stripped one period earlier (dark blue) show a much clearer separation between the fast and slow stars. In this case, the slower stars form an overdensity that includes the slowest stars from the previous period (light blue). Once the peaks have expanded significantly (green and orange periods), the slow and fast stars are no longer clearly separable. The peaks overlap in such a way that the density saturates, resulting in a constant value along the stream. The grey horizontal line in the top panel shows the value at which the density saturates. For an older stream, additional periods will increase the length of the constant density region but not its saturation value. The total length of the stream is determined by the fastest stars stripped during the first pericentre passage (red). In this case, each arm of the stream is about $4.62$~rad long.

In general, streams generated by globular clusters following an eccentric orbit similar to M68 will present symmetric arms with the following density structure:
\begin{enumerate}
    \setlength\itemsep{0.5em}
    \item Large underdensity close to the cluster.
    \item A concentrated overdensity corresponding to the last pericentre passage.
    \item Density oscillations of smaller amplitude, depending on the frequency distribution of the stripped stars.
    \item Constant density of length proportional to the number of pericentre passages undergone by the progenitor.
    \item A long density decline of a length determined by the fastest stars stripped during the first pericentre passage.
\end{enumerate}
In practice, we only observe sections of streams, particularly when they are close to the Sun. Therefore, the section of a stream that is observed determines the research methodology and the possible conclusions that can be obtained. In the following section, we use \textit{Gaia} data to determine the visible part of the M68 stream and its surface density.


\section{GDR3 M68 stream star selection}\label{GDR3_selection}

\begin{figure}
\includegraphics[width=1.0\columnwidth]{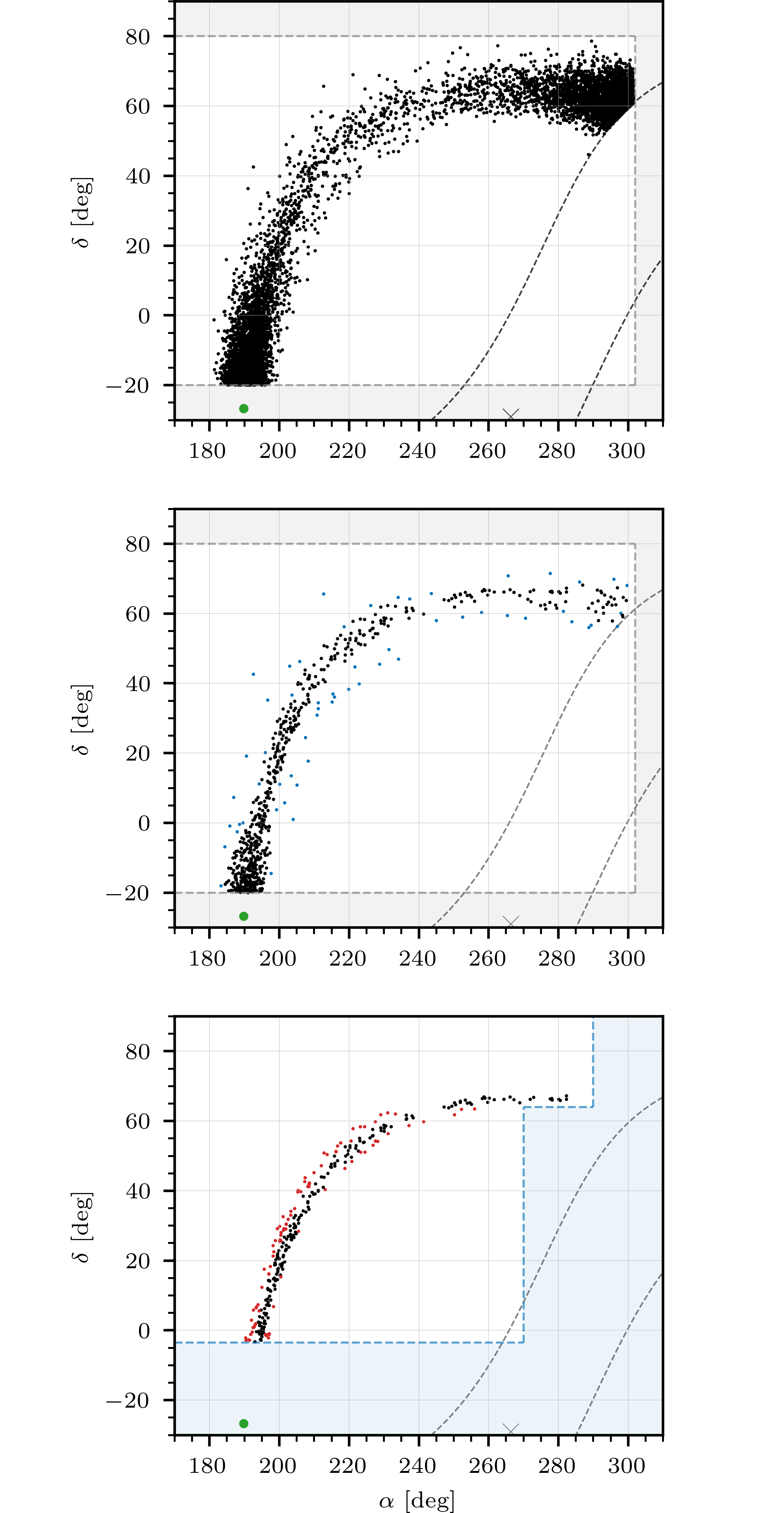}
\caption{Sky coordinates in right ascension $\alpha$ and declination $\delta$. The large green dot indicates the location of the cluster M68, and the grey cross marks the Galactic centre. The limits of the disc at $b=15$~deg are shown as grey dashed lines. \textit{Top:} GDR3 stars that pass the quality cuts and have been pre-selected due to their large intersection with the phase-space volume following the orbit of the cluster. The area excluded by the quality cuts is shaded in grey and its border is highlighted by a grey dashed line. \textit{Middle:} Same as the top panel, but showing only the stars compatible with the CMD of the cluster. The stream stars are shown in black and the isolated stars are shown in blue. \textit{Bottom:} Final GDR3 stream star selection. This is the same as the middle panel, but with the isolated stars and the stars in areas with high foreground contamination removed. These areas are within the blue shaded zone, the limits of which are marked with blue dashed lines. The stars of the main component of the stream are shown in black, while the star in the envelope are shown in red.}
\label{gaia_selection_1}
\end{figure}

In this section, we improve the M68 star selection by \citet{2019MNRAS.488.1535P} using data from the \textit{Gaia} Data Release 3 (GDR3) catalogue \citep{2023A&A...674A...1G}. We apply the following cuts to reduce the amount of data to be analysed and remove low-quality measurements:
\begin{enumerate}
\setlength\itemsep{0.5em}
 \item \texttt{parallax} $<1/0.3$ mas
 \item $-20\leqslant$ \texttt{dec} $\leqslant80$ deg
 \item $170\leqslant$ \texttt{ra} $\leqslant302$ deg
 \item \texttt{b} $>15$ deg
 \item \texttt{bp\_rp} $\neq$ Null
 \item \texttt{ruwe} $<1.2$
 \item \texttt{visibility\_periods\_used} $\geqslant10$
 \item \texttt{duplicate\_source} $=$ False
\end{enumerate}
For each star that passes these cuts, we determine its intersection with a phase-space volume surrounding the orbit of the cluster. This volume is defined by a bundle of orbits computed using different phase-space coordinates of the cluster and different potentials. This method is described in detail in Appendix~C of \citet{2019MNRAS.488.1535P}, and Appendix~\ref{App5} provides the values of the free parameters used in this case. We pre-select the $6614$ stars with the highest intersection and plot their sky coordinates as black dots in the top panel of Figure~\ref{gaia_selection_1}. We observe that the stars are distributed along a path on the sky pointing towards the cluster, which is marked by a large green dot. The star density is significantly higher in the regions where the cluster orbit goes beyond $\Approx5$ kpc from the Sun (Fig.~9, \citetalias{2025MNRAS.539.2718P} and Section~\ref{mock_selection}) and when its projection is close to the disc. We mark the limits of the disc with two dashed grey lines at Galactocentric latitude of $b=\pm15$~deg.

\begin{figure}
\includegraphics[width=1.0\columnwidth]{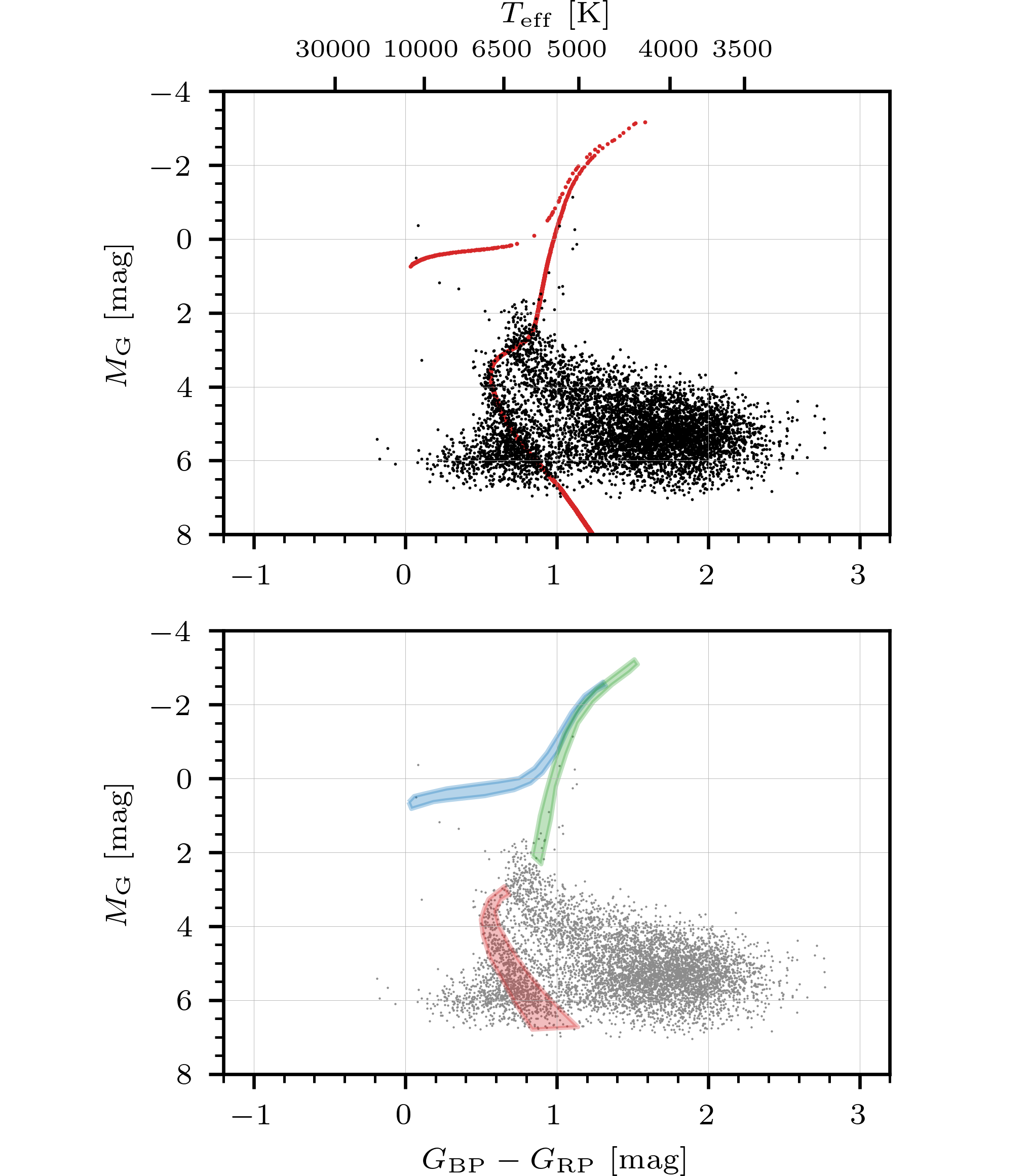}
\caption{Colour Magnitude Diagram (CMD). The absolute magnitude $M_{\rm G}$ is computed using the estimated distance of the stars from the cluster orbit. The colour index $\BPRP$ and the $M_{\rm G}$ include a correction for dust reddening. \textit{Top:} The black dots show the GDR3 stars that pass the quality cuts and have been pre-selected due to their large intersection with the phase-space volume following the orbit of the cluster. A synthetic population of the globular cluster M68, generated by PARSEC/COLIBRI, is shown in red. \textit{Bottom:} Same as top panel, but with the stars shown in grey. The polygons used to define the CMD cuts are shown as coloured areas for the main sequence (red), the red giant branch (green), and the horizontal branch (blue).}
\label{gaia_selection_2}
\end{figure}

To determine the absolute magnitude of the stars, we estimate their distance from the closest point of the orbit of the cluster, as described in \S5.3 of \citetalias{2025MNRAS.539.2718P}. We also apply the dust reddening correction introduced in App.~B of \citetalias{2025MNRAS.539.2718P}. In the top panel of Figure~\ref{gaia_selection_2}, we show the Hertzsprung–Russell or Colour Magnitude Diagram (CMD) of the pre-selected stars. We also include in red a synthetic population of M68 generated by PARSEC/COLIBRI (\S2.1, \citetalias{2025MNRAS.539.2718P}). We observe that most of the stars are distributed in two clumps, both of which have an extension for brighter stars. The smaller clump, located at $0.2\LessSim\BPRP\LessSim1.2$ mag, has an extension at $M_{\rm G}\Approx4$ mag that fits with the main sequence of the M68 globular cluster. Most of these stars belong of the stream. In order to select the stars that are compatible with the CMD of M68, we define three polygons that correspond to the main sequence (red), the red giant branch (green), and the horizontal branch (blue). We have excluded a section of the turn off because it is highly contaminated by stars from the extension of the second clump. The polygons are shown in the bottom panel of Figure~\ref{gaia_selection_2}. The stars located within one of these polygons are selected.

In the middle panel of Figure~\ref{gaia_selection_1}, we plot as small dots the $601$ pre-selected stars compatible with the CMD of M68. The stream clearly emerges as an elongated overdensity compared to the background. We eliminate the $58$ isolated stars from the selection, defined as the stars with fewer than $3$ neighbours within an angular radius of $2.5$~deg. These stars are shown in blue in Figure~\ref{gaia_selection_1}. The sections of the stream that are close to the cluster and close to the disc are highly contaminated by foreground stars. We eliminate these sections by defining the limits of the observable part of the stream:
\begin{enumerate}
\setlength\itemsep{0.5em}
 \item $\delta\,\geqslant-3.5$ deg
 \item $\alpha\leqslant290\,$ for $\,\delta\geqslant64$ deg
 \item $\alpha\leqslant270\,$ for $\,\delta<64$ deg.
\end{enumerate}
These limits are marked as blue dashed lines and define the boundaries of the blue shaded areas in the bottom panel of Figure~\ref{gaia_selection_1}. The sections of the stream beneath these areas are difficult to observe using data from the GDR3 catalogue alone. Conversely, the observable section of the stream is projected onto the white area of the sky. The main peak of the stream, which corresponds to the last pericentre passage, lies below the $\delta=-3.5$~deg limit. Its location and the possibility of observing it are discussed in Appendix~\ref{App6}. After removing the isolated stars and those within the blue areas, we obtain $291$ stream stars in the final GDR3 selection. These stars are plotted as small coloured dots in the bottom panel of Figure~\ref{gaia_selection_1}, where black indicates that the star is consistent with being originated from the globular cluster, and red indicates otherwise (Section~\ref{stream_width}).

The GDR3 catalogue provides radial velocities for $2$ stars in the final selection. We checked the Survey of Surveys (SoS) DR1 compilation \citep{2022A&A...659A..95T}, which includes radial velocities from several sources, and found $2$ star from the LAMOST DR5 catalogue \citep{2012RAA....12..723Z} and $6$ stars from the SEGUE catalogue \citep{2009AJ....137.4377Y}. Additionally, the \texttt{STREAMFINDER} catalogue \citep{2021ApJ...914..123I} includes $11$ stars corresponding to the Fjörm stream\footnote{The M68 stream is named Fjörm and labelled 22 in the catalogue.}. However, we decided not to include radial velocities in the analysis, as all the sources together only provide radial velocities for about $7$ per cent of the final selection. Furthermore, the DESI Milky Way Survey (MWS) \citep{2023ApJ...947...37C} is expected to provide spectroscopic measurements for $\Approx100$ stream stars in future releases of the DESI catalogue. This will significantly improve the separation between stream and foreground stars, as well as enabling a more detailed spectroscopic analysis of the M68 stream.

\subsection{Stream width}\label{stream_width}

\begin{figure}
\includegraphics[width=1.0\columnwidth]{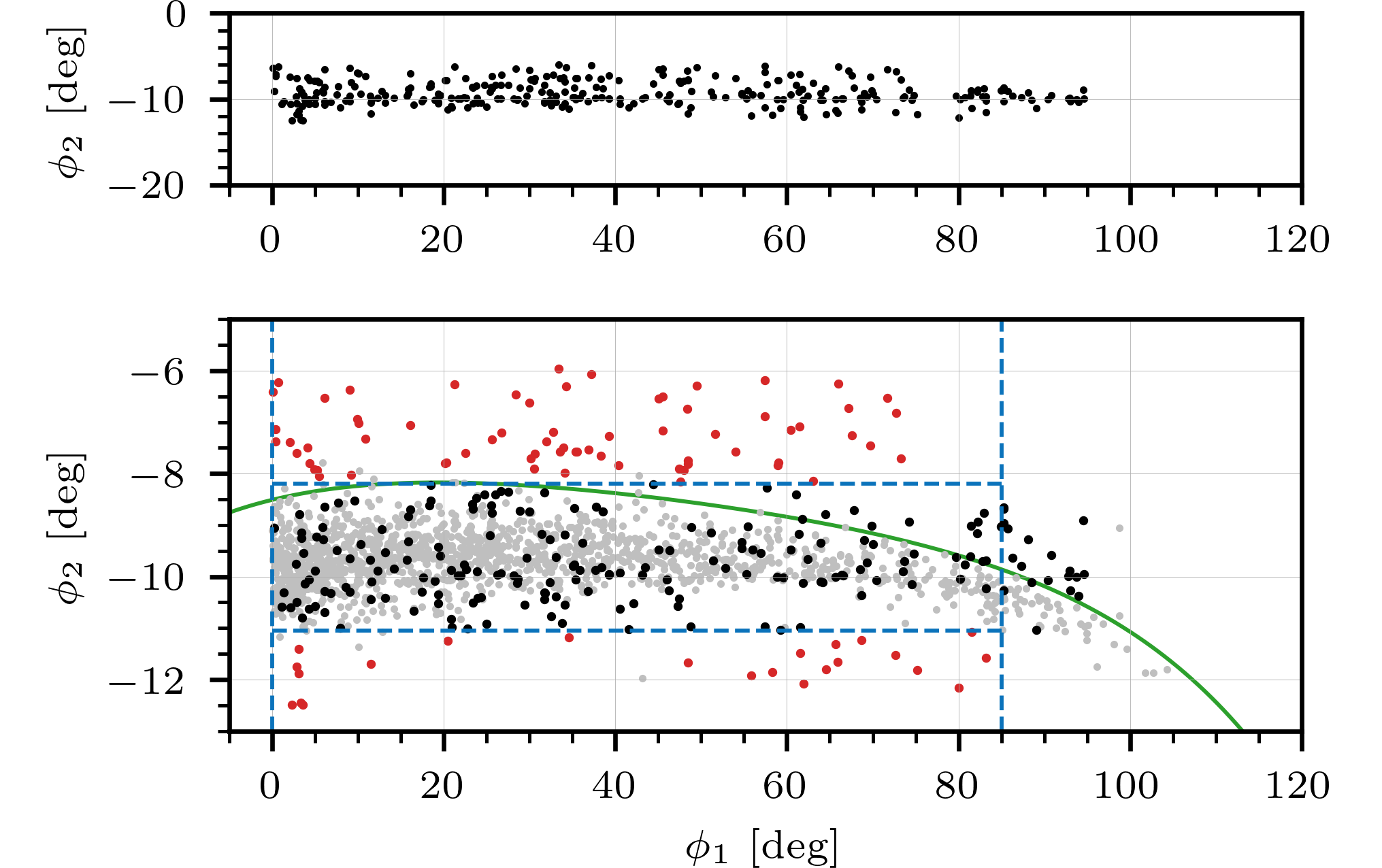}
\caption{Sky coordinates aligned with the stream ($\phi_1,\phi_2$). \textit{Top:} GDR3 stream star selection plot with an aspect ratio of 1:1. \textit{Bottom:} Same as top, but with a plot aspect ratio of 6:1. The stars of the main component of the stream are shown as black dots, and the stars of the envelope are shown as red dots. The boundaries between the components are shown as blue dashed lines. The simulated stream stars from the \nbody\ simulation are shown as small grey dots. The cluster orbit for $t>0$ is shown as a solid green line.}
\label{phi}
\end{figure}

In order to determine the width of the stream, we rotate the sky coordinates to minimise the distortion caused by the projection. In Appendix~\ref{App1} we introduce the rotation matrix from $\{\delta,\alpha\}$ to the coordinates $\{\phi_1,\phi_2\}$, which minimise the curvature of the stream. In the top panel of Figure~\ref{phi}, we plot the rotated sky coordinates of the GDR3 star selection. The width of the stream is approximately $6$~deg and remains constant along the observable section of the stream. Only the part farthest from the globular cluster ($\phi_1\GtrSim80$~deg) has a width smaller than $3$~deg, most likely because this section is the most distant from the Sun and only the brightest stars are visible.

In the bottom panel of Figure~\ref{phi}, we compare the width of the stream with the \nbody\ simulation. The simulated stars are plotted as grey dots and the GDR3 star selection is plotted as coloured dots. We observe that the simulation is approximately contained within the limits of $-11\LessSim\phi_2\LessSim-8$~deg, while the observed stream is about twice as wide spanning about $-12\LessSim\phi_2\LessSim-6$~deg. Additionally, the simulated leading arm of the stream extends beneath the cluster orbit, which is shown as a solid green line in the bottom panel of Figure~\ref{phi}. Approximately half of the width of the stream is located above the orbit. We note that the orbit depends on the Milky Way potential and that its exact location is especially sensitive for $\phi_1\GtrSim80$~deg. However, variations in the potential do not significantly affect the position of the orbit close to the cluster, for $\phi_1\Approx0$~deg.

We identify the stars that can be traced back to the globular cluster, and define them as the main component of the stream. The remaining stars are defined as members of the envelope. The stars in the main component are those that fulfil one of the following conditions:
\begin{enumerate}
\setlength\itemsep{0.5em}
 \item $0\leqslant\phi_1\leqslant85$ deg and $-11.5\leqslant\phi_2\leqslant-8.2$ deg
 \item $\phi_1>85$ deg
\end{enumerate}
We do not apply the constraint on $\phi_2$ in the second condition because, in general, a long stream presents curvature in the rotated coordinate frame, as can be seen in the \nbody\ simulation (grey dots) in the bottom panel of Figure~\ref{phi}. In the same panel, the $199$ stars of the GDR3 selection that fulfil the above conditions are shown as black dots, and the $92$ stars of the envelope are shown as red dots. The sky coordinates of these stars in the ICRS reference frame are also shown in the bottom panel of Figure~\ref{gaia_selection_1} with the same colour code. Most of the envelope stars are located above the stream ($\phi_2>-8.2$~deg) and are distributed uniformly along it. A small fraction is located below the stream, close to the cluster, but mainly within the range $60\LessSim\phi_1\LessSim90$.

\begin{figure}
\includegraphics[width=1.0\columnwidth]{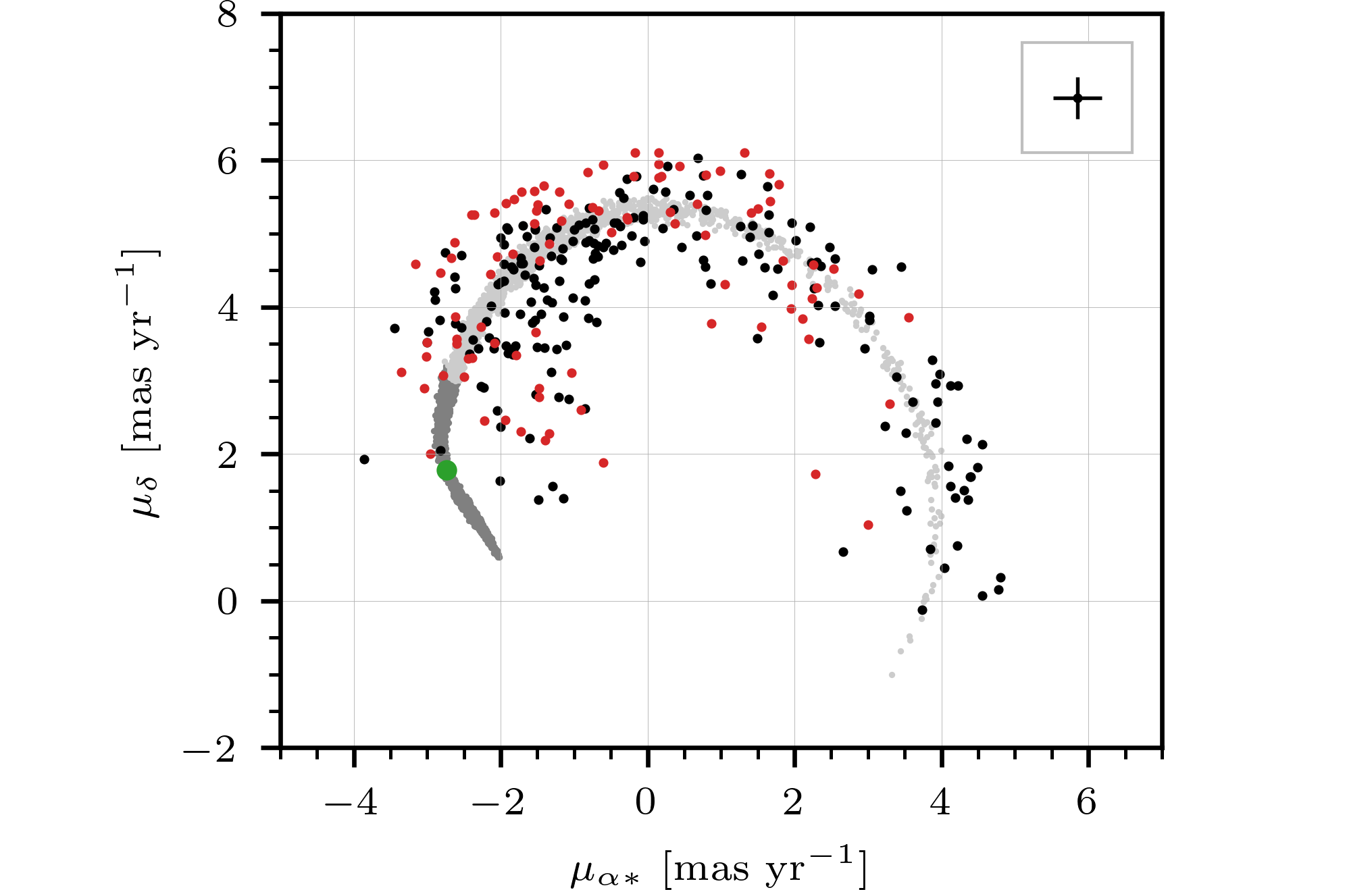}
\caption{Proper motions of the GDR3 star selection. The stars of the main component are shown as black dots and those of the envelope as red dots. The observable section of the \nbody\ simulation is shown as small light grey dots, and the rest of the simulated stream as small dark grey dots. The mean of the observed proper motions of the M68 cluster is shown as a large green dot. In the top right floating panel, we illustrate the size of the observational uncertainties using a star with a \textit{Gaia} magnitude of $G\simeq19.17$~mag, which is equal to the mean magnitude of the selected stars.}
\label{gaia_selection_pm}
\end{figure}

In Figure~\ref{gaia_selection_pm}, we plot the proper motions of the GDR3 star selection. The stars of the main component (black dots) fit with the \nbody\ simulation (grey dots). The larger dispersion is explained by the observational uncertainties. In the top right corner, we illustrate the size of the uncertainties using a star with a \textit{Gaia} magnitude of $G\simeq19.17$~mag, which is equal to the mean magnitude of the selected stars. There is no a clear distinction between the main component and the envelope (red dots). Nevertheless, the envelope has systematically larger proper motions in the range $-2\LessSim \mu_{\alpha\ast} \LessSim 2$~mas~yr$^{-1}$ than the main component of the stream. This suggests that the stars in the envelope may be slightly closer to the Sun than the main component.

Other streams present a similar structure consisting of a main component and a fainter envelope. For example Jhelum \citep{2023A&A...669A.102W, 2024A&A...683A..14A} or GD-1 \citep{2019ApJ...881..106M, 2025ApJ...980...71V, 2025ApJ...988...45T}. In the case of M68, the size of the globular cluster is well constrained (\S2.2, \citetalias{2025MNRAS.539.2718P}), and variations within the observational uncertainties do not significantly increase the width of the simulated stream. Therefore, we present a list of possible causes that could explain this discrepancy in width, as well as similar features of other streams:
\begin{enumerate}
\setlength\itemsep{0.5em}

 \item Prior to accretion, the progenitor cluster orbited a Milky Way satellite galaxy, generating an stellar stream due to the tidal forces of the satellite. After accretion, this stream is mixed with a new one generated by the cluster due to the tidal forces of its new host galaxy \citep{2020ApJ...889..107C, 2021MNRAS.501..179M, 2022MNRAS.511.2339Q}.

 \item A cluster close to a chaotic phase-space region or resonant trapped orbit family can develop a stream with a wide range of star frequencies. This causes the width of the stream to expand, forming a fan-shaped feature. \citep{2021MNRAS.501.1791Y, 2023ApJ...954..215Y}.

 \item Perturbations from a tilting disc can produce widening of stellar streams \citep{2024ApJ...969...55N}.

 \item Interactions with Milky Way satellites, such as the Sagittarius dwarf galaxy, the LMC \citep[e.g.][]{2022MNRAS.510.2437E, 2022MNRAS.516.1685D} or the Galactic bar \citep[e.g.][]{2016ApJ...824..104P, 2023A&A...678A.180T} could alter the structure of the stream and explain the observed widening.

 \item According to simulations, interacting with hypothetical dark matter subhaloes can increase the width of the stream \citep[e.g.][]{2023ApJ...953...99C, 2024ApJ...975..135C}.
\end{enumerate}

Another explanation is that the M68 stream comprises several star populations of different origins. In fact, the metallicity of the stream is broader than that of similar globular cluster streams, such as Palomar 5 \citep{2022MNRAS.516.5331M}. Therefore, it is possible that some stream stars, especially those in the envelope, are remnants of an accreted system that included M68 and was formed from a dwarf galaxy and other globular clusters (Section~\ref{comp_pre_est}). This question remains open and will require more precise and complete spectroscopic surveys in future studies.

\subsection{Stream surface density}\label{str_surface_density}

As the observed width of the stream cannot be explained by an axisymmetric potential, such as that used in the \nbody\ simulation, and given the possibility that the stream consists of multiple components, we restrict our analysis to the stars from the main component of the stream. These stars are shown as black dots in the bottom panels of Figure~\ref{gaia_selection_1}, \ref{phi}, and \ref{gaia_selection_pm}. To eliminate projection distortions, especially prevalent for long streams, we estimate the angle along the principal axis of the stream relative to the globular cluster $\Delta\bar{\theta}_1$ for each stream star. This angle is estimated from the closest point of the cluster orbit, using the same method as that used to estimate the Heliocentric distances (\S5.3, \citetalias{2025MNRAS.539.2718P}). There is a systematic offset of $5.47$~mrad between the exact and the estimated values of $\Delta\bar{\theta}_1$ in the observable section of the stream. After the correction, the standard deviation of the error is $5.12$~mrad, which we consider to be negligible.

\begin{figure}
\includegraphics[width=1.0\columnwidth]{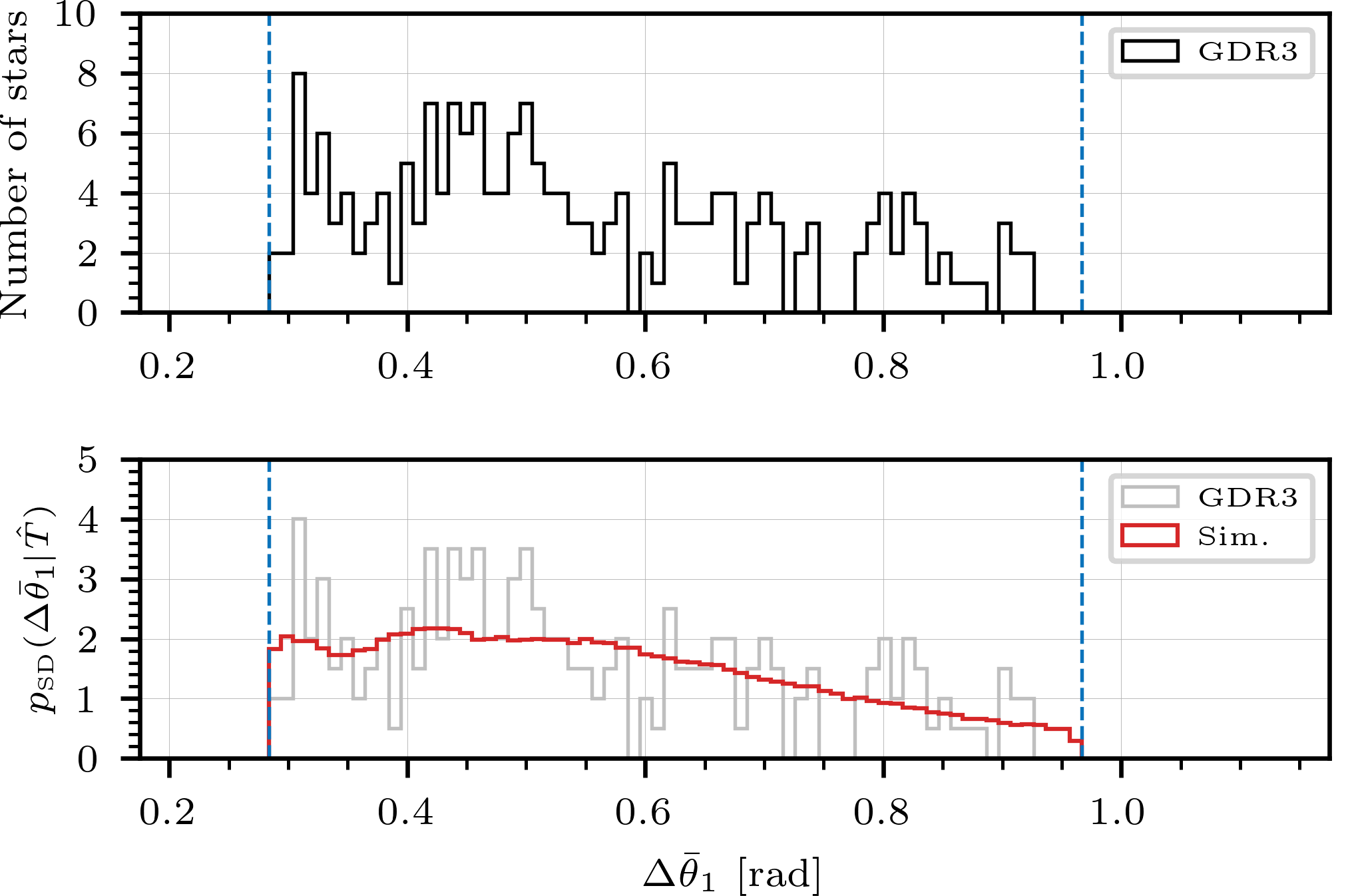}
\caption{Surface density of the main component of the stream as a function of the angle along the principal axis of the stream relative to the globular cluster $\Delta\bar{\theta}_1$. The histogram has bins of approximately $10$~mrad. The blue dashed lines indicate the boundaries of the observable section of the stream. \textit{Top:} GDR3 stream star selection. \textit{Bottom:} Same as top panel but normalised (grey). Best-fitting simulated surface density $p_{\SD}$ corresponding to a stream of age $T=3039.7$~Myr (red).}
\label{surface_density}
\end{figure}

In the top panel of Figure~\ref{surface_density}, we plot the observed surface density of the main component of the stream, estimated using a histogram with bins of approximately $10$~mrad. The limits of the observable section $\Delta\bar{\theta}_1\simeq\Range{0.284}{0.967}$~rad are shown as blue dashed vertical lines. The surface density presents a double-peak structure close to the cluster. It is the highest at $\Delta\bar{\theta}_1\Approx0.3$~rad, and peaks at $\Delta\bar{\theta}_1\Approx0.45$~rad, which corresponds to the point of closest approach to the Sun. For larger angles, the observed surface density declines as the stream moves away from the Sun.

\section{Mock star selection}\label{mock_selection}

In this section, we introduce the process of generating mock final selections of stream stars. These mocks are used to determine the PDF of the surface density of the stream, taking into account the effect of the selection function, observational uncertainties and the process of separating the stream stars from the foreground. Additionally, we use a mock random sample of observed stream stars to test the methods for determining the age and mass loss of the progenitor cluster. This process consists of ten steps:
\begin{enumerate}
\setlength\itemsep{0.5em}
 \item[{\crtcrossreflabel{(1)}[item_sim]}] Simulation of the stellar stream of age $T$ in angle-frequency space (Section~\ref{model}).

 \item[{\crtcrossreflabel{(2)}[2]}] Conversion from the angle-action coordinates to Galactocentric cartesian coordinates using the Torus Mapper code (Subsection~\ref{actions_inv_trans}).

 \item[{\crtcrossreflabel{(3)}[3]}] Estimation of the angle $\Delta\bar{\theta}_1$ from the cluster orbit (Subsection~\ref{str_surface_density}).

 \item[{\crtcrossreflabel{(4)}[4]}] Generation of a synthetic stellar population compatible with the globular cluster (\S2.1, \citetalias{2025MNRAS.539.2718P}).

 \item[{\crtcrossreflabel{(5)}[5]}] Application of the dust reddening extinction (App.~B, \citetalias{2025MNRAS.539.2718P}).

 \item[{\crtcrossreflabel{(6)}[item_sel_func]}] Application of the GDR3 selection function (\S5.1, \citetalias{2025MNRAS.539.2718P}).

 \item[{\crtcrossreflabel{(7)}[7]}] Estimation of the GDR3 astrometric, photometric, and spectroscopic observational uncertainties (\S5.2, \citetalias{2025MNRAS.539.2718P}).

 \item[{\crtcrossreflabel{(8)}[8]}] Generation of random samples following Gaussian distributions given by the simulated values and the observational uncertainties (\S5.4, \citetalias{2025MNRAS.539.2718P}).

 \item[{\crtcrossreflabel{(9)}[item_sel]}] Selection process (Section~\ref{GDR3_selection}).

 \item[{\crtcrossreflabel{(10)}[item_main_comp]}] Determination of the stars of the main component of the stream (Subsection~\ref{stream_width}).

\end{enumerate}
The following summary outlines the process of selecting stream stars from the GDR3 catalogue \ref{item_sel} introduced in Section~\ref{GDR3_selection}:
\begin{enumerate}
\setlength\itemsep{0.5em}
 \item[{\crtcrossreflabel{(9.1)}[91]}] Application of the cuts to reduce the total amount of data. The quality cuts in \texttt{ruwe}, \texttt{visibility\_periods\_used}, and \texttt{duplicate\_source} are not included in this simulation.

 \item[{\crtcrossreflabel{(9.2)}[92]}] Pre-selection of stars compatible with a phase-space volume following the orbit of the cluster.

 \item[{\crtcrossreflabel{(9.3)}[93]}] Estimation of the distance of the stars from the cluster orbit.

 \item[{\crtcrossreflabel{(9.4)}[94]}] Selection of stars compatible with the CMD of the progenitor cluster.

 \item[{\crtcrossreflabel{(9.5)}[]}] Elimination of the sections of the stream obscured by foreground contamination.

\end{enumerate}

\begin{figure*}
\includegraphics[width=1.0\textwidth]{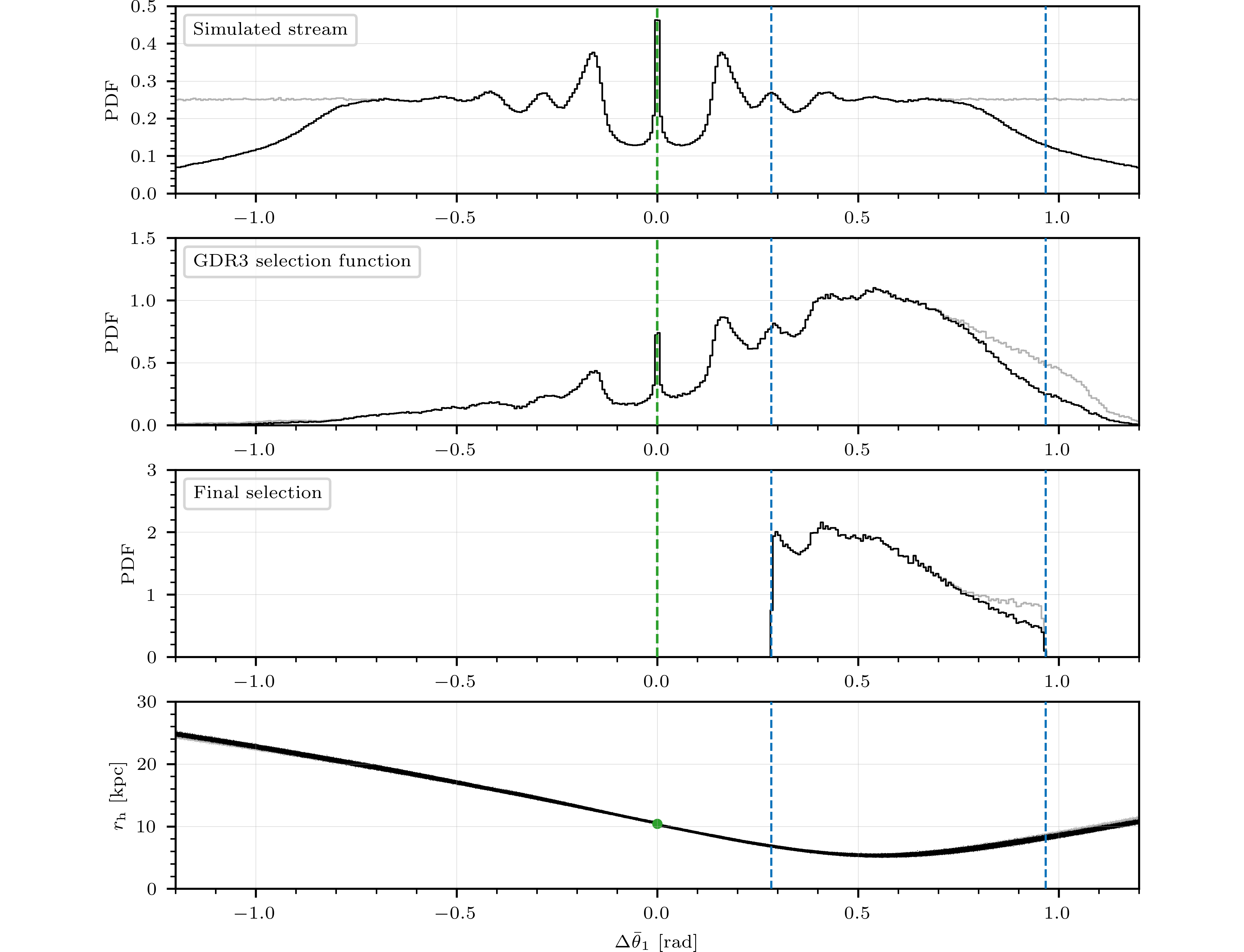}
\caption{PDF of the surface densities as a function of the angle along the principal axis of the stream $\Delta\bar{\theta}_1$. The surface densities are estimated using histograms of $6.25$~mrad bins. The black histogram corresponds to a stream of age $T=3039.7$~Myr. The grey histogram corresponds to a stream that has reached saturation for the entire depicted angular range. Its area has been scaled to enable comparison with the black histogram. The cluster position is indicated by a dashed green line, and the limits of the observable section are indicated by dashed blue lines. \textit{Top:} Simulated streams. \textit{Middle-top:} Stars from the simulated streams that pass the GDR3 selection function.  \textit{Middle-bottom:} Stars from the main components of the simulated streams that pass the GDR3 selection function and the selection process. \textit{Bottom:} Heliocentric distance $r_{\rm h}$ of the stream stars as a function of the angle along the principal axis of the stream $\Delta\bar{\theta}_1$. In this case, the position of the globular cluster is indicated with a large green dot.}
\label{surface_density_model}
\end{figure*}

Figure~\ref{surface_density_model} illustrates several steps of this process. In the top panel, we plot in black the surface density of both arms of a simulated stream of age $T=3039.7$~Myr \ref{item_sim}. This stream corresponds to the configuration that best fits the GDR3 star selection, as discussed in Section~\ref{estimations}. The surface density is estimated from a random sample of approximately $1.3\pd{7}$ stars using a histogram of $6.25$~mrad bins, and is expressed as a function of the angle along the principal axis of the stream $\Delta\bar{\theta}_1$. The position of the globular cluster is indicated by a green dashed line and the boundaries of the observable section are marked by blue dashed lines. This stream has symmetric arms and four well-defined peaks, with the saturation density covering approximately two-thirds of the observable section. For reference, the surface density of a simulated stream that has reached saturation for the entire depicted angular range is shown as a grey histogram.

In the middle-top panel of Figure~\ref{surface_density_model}, we plot the stars that pass the GDR3 selection function \ref{item_sel_func}. This function depends on the sky coordinates and the $G$ magnitude of the stars. The determining factor is the distance. In the bottom panel of Figure~\ref{surface_density_model}, we plot the Heliocentric distance of the stream stars as a function of $\Delta\bar{\theta}_1$, and mark the cluster position at $r_{\rm h}=10.404$~kpc with a large green dot. The entire trailing arm of the stream lies beyond $10$~kpc, so only the brightest stars are observable. Conversely, the leading arm is closer to the Sun, at up to $\Approx5$~kpc. The surface density appears as a wide peak within the interval $\Delta\bar{\theta}_1\Approx\Range{0.4}{0.65}$~rad. For $\Delta\bar{\theta}_1\Approx0.35,\ {\rm and}\ 0.5$~rad, the underdensities between the real peaks are visible. For values $\Delta\bar{\theta}_1\GtrSim0.65$~rad, the density declines due to the increasing distance of the stream stars.

In the middle-bottom panel of Figure~\ref{surface_density_model}, we plot the stars that pass the selection process \ref{item_sel} and belong to the main component of the stream \ref{item_main_comp}. Approximately $2.6\pd{5}$ stars pass through these cuts. This number is large enough to account for the full distribution of observational uncertainties of the stream stars, and therefore ensures that the surface density has converged to its true value. The mock observed distribution peaks at $\Delta\bar{\theta}_1\GtrSim0.4$~rad. This peak does not correspond to the angle of closest approach of the stream, which is $\Delta\bar{\theta}_1\simeq0.56$~rad. It is possible that some stars around $\Delta\bar{\theta}_1\Approx0.56$~rad are lost during the selection process. In particular, stars with large proper motions uncertainties have a smaller intersection with the volume following the cluster orbit. For angles $\Delta\bar{\theta}_1\GtrSim0.55$~rad, the density declines almost linearly up to the limit of the observable section.

\section{Age and mass loss estimation of the M68 stellar stream}\label{estimations}

We estimate the age $T$ of the M68 stellar stream, or equivalently the accretion time of the cluster, using the posterior distribution $p\var{T|d}$ given by the Bayes' theorem:
\begin{equation}\label{bayes}
p\var{T|\:\!d} = \frac{\mathcal{L}\var{d\:\!|T} \, p\var{T}}{p\var{d\:\!}},
\end{equation}
where $d$ is the GDR3 selection of observed stars obtained in Section~\ref{GDR3_selection} and plotted in the top panel of Figure~\ref{surface_density}. The likelihood function $\mathcal{L}\var{d\:\!|T}$ is defined as:
\begin{equation}\label{likelihood}
\mathcal{L}\var{d\:\!|T} \equiv \prod_{\Delta\bar{\theta}_1 \in d} \, p_{\SD}\var{\Delta\bar{\theta}_1|T},
\end{equation}
where the product is over all elements of the dataset $d$ and $p_{\SD}\var{\Delta\bar{\theta}_1|T}$ is the PDF of the surface density within the observable limits. This distribution is estimated by combining $10^3$ samples of mock observed stream stars, generated following the procedure described in Section~\ref{mock_selection}. In total, the samples contain a number of stars sufficiently large to cover the distributions of observational uncertainties completely, and therefore ensure that $p_{\SD}$ has converged to the true probability distribution. We estimate the PDF from the samples using a histogram with $150$ bins within the observed range $\Delta\bar{\theta}_1\in\Range{0.284}{0.967}$~rad, which corresponds to bins of approximately $4.55$~mrad. We verified that this number of bins yields the optimal fit by taking a quarter of the star sample to generate the histogram and using the remaining three quarters to determine the best fit.

The prior distribution $p\var{T}$ is taken as a uniform distribution defined from the present time $T=0$ to $T=12$~Gyr, which is the mean of the estimated age of the cluster (Table~1, \citetalias{2025MNRAS.539.2718P}). The best-fitting configuration $\hat{T}$ is obtained by maximising the posterior distribution:
\begin{equation}\label{max_dkl}
\hat{T} \equiv \underset{T}{\mathrm{argmax}}\:\! \big( p\var{T|\:\!d} \big).
\end{equation}
This methodology is tested in Appendix~\ref{App7} using the \nbody\ simulation. In this case, we recover the age of the stream $T$ with an error of approximately $2$~Myr, and estimate the measurement uncertainty to be approximately $10$ per cent of the best-fitting value.

\begin{figure}
\includegraphics[width=1.0\columnwidth]{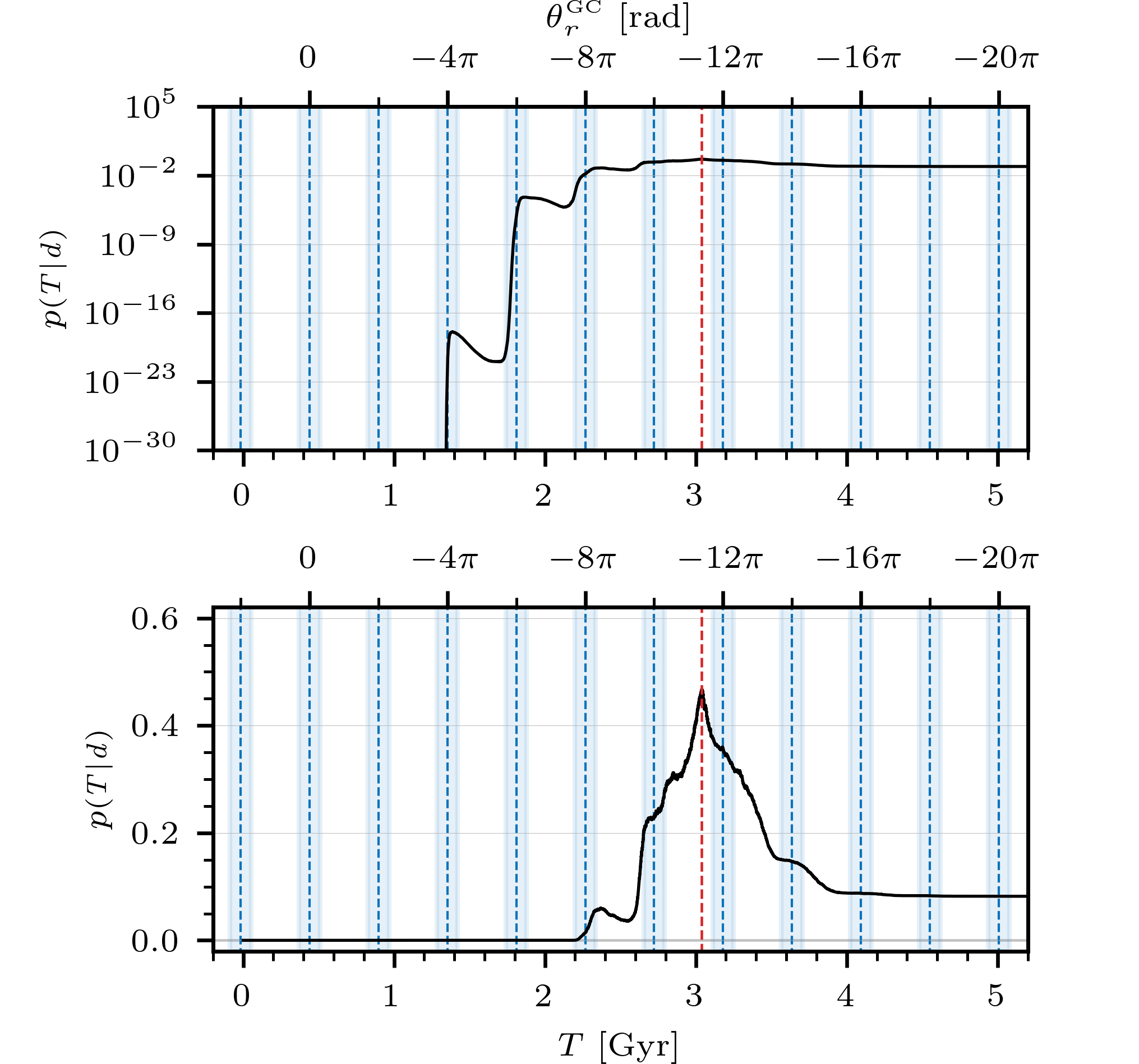}
\caption{Posterior distribution $p$ as a function of the age of the stream $T$, computed for the GDR3 star selection dataset $d$. Each pericentre passage is marked with a blue dashed line, and the interval centred at the pericentre with a width of $\pm70$~Myr is shaded in blue. The maximum of the posterior distribution $\hat{T}$ is marked with a red dashed line. The upper axis shows the radial angle of the globular cluster $\theta_r^{\GC}$. \textit{Top:} Logarithmic scale. \textit{Bottom:} Linear scale.}
\label{posterior}
\end{figure}

Figure~\ref{posterior} shows the posterior distribution as a function of the stream age $T$. The radial angle of the globular cluster $\theta_r^{\GC}$ is also indicated on the upper ordinate axis. Each pericentre passage is marked with a blue dashed line. An interval of $\pm70$~Myr centred at the pericentres is shaded in blue to represent the approximate size of the peak. In the top panel, we plot the posterior distribution using a logarithmic scale. Moving from younger to older ages, we can see how the posterior distribution increases with each pericentre passage, reaching a maximum at $\hat{T}=3039.7$~Myr. This value corresponds to the best-fitting configuration, and is marked in the figure with a red dashed line. This behaviour is explained by the fact that the stream grows in bursts caused by tidal shocks at each pericentre passage, rather than linearly over time. For $T<\hat{T}$, the value of the posterior distribution decreases between the peaks because stars released after the pericentre passage move slower and only populate the central part of the observable section of the stream. Consequently, the central density of the stream increases, but its length remains unchanged. This worsens the fit because the observed stream is longer than the simulated one.

We also show the posterior distribution using a linear scale in the bottom panel of Figure~\ref{posterior}. After the maximum $\hat{T}$, the posterior distribution decreases for the subsequent two pericentre passages. These two additional tidal shocks populate the region $\Delta\bar{\theta}_1\GtrSim0.8$~rad, which has low surface density in the observed data, as shown in the top panel of Figure~\ref{surface_density}. For $T\GtrSim4$~Gyr, the posterior distribution remains constant. This is explained because the stream reaches saturation within the observable interval. For older streams, the density within the observable section does not change, and therefore the posterior distribution has a constant value.

We plot the PDF of the surface density for the best-fitting configuration $p_{\SD}\var{\Delta\bar{\theta}_1|\hat{T}}$ in red in the bottom panel of Figure~\ref{surface_density}, and compare it to the distribution of GDR3 selected stream stars, shown in grey. This distribution accurately describes several features of the data, including the presence of two peaks and the small underdensity between them, the length of the central peak, and the linear decay at larger angles. The main discrepancy is the excess of observed stars in the range $\Delta\bar{\theta}_1\Approx\Range{0.4}{0.5}$~rad. This can be explained by a number of envelope stars being projected within the limits of the main component, by the inclusion of foreground stars, or by statistical biases present in the observed sample. Accurate modelling of these density fluctuations is necessary in order to estimate the age of the stream correctly. In Appendix~\ref{App10}, we demonstrate that simpler models assuming constant mass loss and a constant frequency distribution for the stripped stars cannot reproduce these density variations, and consequently bias the estimated age of the stream. Nevertheless, we consider the discrepancy between our model and the observed sample to be small, and we conclude that the best-fitting configuration is a good representation of the data. The PDF of the best-fitting configuration is also shown in Figure~\ref{surface_density_model} as a black histogram.

We estimate the age of the simulated stream to be $T=\round[2]{3.0397}_{-\round[2]{0.28704979}}^{+\round[2]{5.63203224}}$~Gyr, where the uncertainties are calculated using the methodology introduced in Appendix~\ref{App8}. The lower uncertainty is small because the stream cannot be shorter than the observed GDR3 star sample. Conversely, the upper uncertainty is large. This is due to the small difference between a stream of $T=3039.7$~Myr and a stream that has reached saturation. This difference can be seen in the middle-bottom panel of Figure~\ref{surface_density_model} for angles $\Delta\bar{\theta}_1\GtrSim0.8$~rad, where the saturated stream is shown as a grey histogram. In this region, it is difficult to determine the surface density with high precision in practice. This is because the stream is located at a distance of $\Approx8$~kpc from the Sun. At this distance, the selection procedure can erroneously exclude stars with large proper motion errors. Furthermore, once it has reached saturation within the observable section, the surface density is insensitive to the age of the stream. These two phenomena explain the large upper uncertainty of our estimate.

In order to determine the mass loss of the cluster, we adjust the parameter $N_T$ (Section~\ref{num_stripp_stars}) so that the mean of the mock observed star samples (Section~\ref{mock_selection}) equals the number of stars in the observed sample. We note that the model used to generate the mock star samples assumes that the cluster mass and density distribution remain constant throughout its evolution (Section~\ref{num_stripp_stars}). Therefore, we determine $N_T$ assuming that it is independent of the properties of the cluster. We tested this methodology in Appendix~\ref{App7} using the \nbody\ simulation, obtaining uncertainties of about $10$ per cent with respect to the true value. In the GDR3 star selection, there are $199$ stars in the main component of the stream (Section~\ref{GDR3_selection}). Adjusting for this number, we obtain $N_T \simeq 1.718\pm0.104$~stars~Myr$^{-1}$~arm$^{-1}$. Multiplying this by the mean of the Kroupa distribution (Section~\ref{num_stripp_stars}), we obtain a mass loss of approximately $0.496\pm0.030$~M$_{\Sun}$~Myr$^{-1}$~arm$^{-1}$. This implies a total mass lost of $3016.5$~M$_{\Sun}$ since accretion, corresponding to approximately $2.4$ per cent of the cluster initial mass. These values are sufficiently small to be compatible with the model assumption that the mass of the cluster is constant.

\subsection{Comparison to previous studies}\label{comp_pre_est}

The age and mass loss of M68 have been studied by \citet{2025ApJ...980L..18C}, using the Fjörm stream\footnote{The M68 stream is also named Fjörm.} from the \texttt{STREAMFINDER} catalogue \citep{2024ApJ...967...89I}. They estimate $T=2.34\pm0.82$~Gyr, $N_T=1.7\pm0.5$~stars~Myr$^{-1}$~arm$^{-1}$, and mass loss equal to $0.5\pm0.13$~M$_{\Sun}$~Myr$^{-1}$~arm$^{-1}$. These results are consistent with ours, with the estimated age being at the lower end of our uncertainty range. In Appendix~\ref{App9}, we discuss whether this age discrepancy could be explained by excluding the stars in the envelope of the stream. We find that including all stars in the selection reduces the estimated age of the stream, which partially explains the discrepancy in the results.

Our estimate of the age indicates that the globular cluster M68 was accreted about $T\GtrApprox3$~Gyr ago. This is broadly consistent with several major accretion events. Some studies have associated M68 with the Canis Major dwarf \citep{2019MNRAS.486.3180K}, Helmi streams \citep[e.g.][]{2019A&A...630L...4M, 2020MNRAS.498.2472K, 2022MNRAS.513.4107C}, Sagittarius dwarf \citet{2021ApJ...909L..26B}, and Cetus Polar stream \citep{2022ApJ...926..107M}. None of these associations are conclusive. However, according to our best-fitting result, it is  possible that M68 was accreted more recently. As suggested by \citet{2023RAA....23a5013S}, it is possible that M68 is associated with a minor accretion event. Furthermore, integrating the cluster orbit backwards in time over a period $\hat{T}$ reveals that the cluster was accreted at a radius $R\simeq27.6$~kpc, which is close to the apocentre $R_{\rm apo}\simeq31.8$~kpc. This is consistent with the cluster having being accreted. However, the potential has not been optimised for this particular stream. Therefore, we postpone drawing any conclusions until a more realistic potential is available.

\section{Conclusion}\label{conclusion}

The dynamics of a stream is greatly simplified in angle-action coordinates. This allows us to simulate a stream with a high degree of accuracy. The developed model focuses on reproducing the effects of variable tidal forces acting on the cluster as it moves along an eccentric orbit. We find that the mass loss of the cluster, as well as the mean and standard deviation of the frequencies of the stripped stars, can be modelled using a simple double exponential model. This model has four free parameters for each variable and depends on the cluster radial angle. Due to its simplicity, the double exponential opens up the possibility of explaining the perturbations observed in the \nbody\ simulation using a semi-analytical approach. Once the free parameters are calibrated, the model can reproduce the internal structure of a stream obtained with the \nbody\ simulation. This model is an improvement on previous methodologies for simulating streams in angle-action coordinates, which neglected the time dependency of mass loss and stripping frequencies \citep{2014MNRAS.443..423S, 2014ApJ...795...95B}.

This model incorporates several simplifications. The most significant of these, especially for long streams, is the assumption that the cluster mass and its density distribution are constant. Modelling this aspect is crucial for old streams, such as Palomar~5, and for predicting the dissolution time of the progenitor clusters. Furthermore, the mass lost by ejection and evaporation have been neglected. Although this is expected to be less than the mass lost by tidal shocks, it is not negligible throughout the life of a globular cluster. Additionally, the frequencies perpendicular to the principal axis of the stream of the stripped stars are considered to be independent of the radial angle of the cluster. Their correlations with the frequency along the stream have also been neglected. Eliminating these assumptions could marginally improve the accuracy of the simulated surface density.

This model depends on several free parameters. The total number can be reduced by making additional assumptions. For example, the distribution of stripping angles of the stream stars has a minimal impact on the morphology of the stream and can therefore be simplified. In addition, the peak in frequency dispersion of the stripped stars can assumed to be symmetric. In principle, all of these free parameters can be expressed as a function of the properties of the cluster, such as its mass and density distribution, as well as the potential of the Galaxy. Further developments of this model will include such relations.

This model can be used to study the surface density of a stellar stream along its principal axis and its dynamical evolution. The main difficulty in applying this method to study real streams lies in calculating the transformation between cartesian and angle-action coordinates. This is only possible for static potentials, and it is computationally expensive, which limits its range of applicability. The transformation is also less accurate for orbits within strong resonant trapped families or close to these resonances. This model can be used to study stellar streams with a known progenitor similar to globular clusters. Examples include the streams of Palomar~5, NGC~3201, M2, and M5. The first application is the study of the stream generated by the globular cluster M68.

Using data from the GDR3 catalogue, we improved upon previous star selections of the M68 stream, obtaining $291$ stars that are likely to be members of the stream. The width of the stream is too large to be explained only by stars stripped from the cluster. We present several explanations for this phenomenon, including the effects of complex and evolving potentials, or the presence of remnants from an accretion event. This feature can potentially introduce biases in the constraints on the Milky Way potential, particularly for methods that use a simulation of the stream based on the properties of the cluster, or based on integrating stars backwards to bring them back to their progenitor. In addition, the observed surface density of the stream presents several density variations in the region closest to the cluster. These variations may be due to the small size of the star sample. In general, they cannot be reproduced by a model that assumes a constant mass loss and a constant frequency distribution for the stripped stars. If these density variations are real, modelling the internal structure of the M68 stream is necessary to correctly estimate its age and the mass loss of the cluster.

We use an static and axisymmetric potential model for the Milky Way. This potential has not been optimised to fit the M68 stream. Nevertheless, the discrepancies between the simulation and the observations are of about one degree in declination within the observable section of the stream. These discrepancies do not affect the estimation of the surface density. By selecting stars consistent with having been stripped from the cluster, we determine the age of the stream, obtaining $T=\round[2]{3.0397}_{-\round[2]{0.28704979}}^{+\round[2]{5.63197931}}$~Gyr. The uncertainties are dominated by the fact that we only observe a section of the stream. As a consequence that the surface density reaches a saturation level within the observable section and therefore the stream only grows in length, we cannot accurately estimate the upper limit of its age. Conversely, we can precisely estimate the lower limit because the stream cannot be shorter than the observed section. The exact value of the best-fitting configuration depends on the details of the observed surface density. In particular, it depends on the section of the stream that is farthest from the globular cluster, as well as on the separation between the main component and the envelope. Improving the observed star selection could potentially reduce the uncertainty in this estimate. Our current best-fitting age of $T=3039.7$~Myr suggests that M68 was accreted more recently than the major accretion events that dominated the formation of the Milky Way.

We also estimate the number of stripped stars $N_T \simeq 1.717\pm0.104$~stars~Myr$^{-1}$~arm$^{-1}$ and the mass loss to be $0.496\pm0.030$~M$_{\Sun}$~Myr$^{-1}$~arm$^{-1}$. These values are consistent with the modelling assumption that the mass of the cluster is constant. While these results are compatible with previous research, they are approximately $20$ per cent lower than those obtained using the \nbody\ simulation. This suggests an inconsistency between the observed properties of the cluster and those inferred from the stream. Alternatively, this could indicate that a significant number of stars have been lost during the selection process and the determination of the main component of the stream. Further research is required to address this discrepancy.


\section*{Acknowledgements}

This work is supported by NSFC (12573022, 12273021), National Key R\&D Program of China (2023YFA1607800, 2023YFA1607801, 2023YFA1605600, 2023YFA1605601), 111 project (No. B20019), and the science research grants from the China Manned Space Project (No. CMS-CSST-2021-A03). We also acknowledge the Yangyang Development Fund. The computation of this work is done on the Gravity supercomputer at the Department of Astronomy, Shanghai Jiao Tong University.


\section*{Data Availability}
A presentation summarising the content of this paper is available on Zenodo \url{https://zenodo.org/records/17330448}. The Python package \texttt{invi} containing the codes used in this paper is available on GitHub: \url{https://github.com/cgpalau-astro/invi}. The following files related to the M68 stellar stream are available on Zenodo \url{https://zenodo.org/records/17020518}:
\begin{enumerate}
 \setlength\itemsep{0.35em}
 \item \textit{N}\!\!\:-body simulation $T=1500$~Myr
 \item \textit{N}\!\!\:-body simulation $T=3040$~Myr
 \item Mock \textit{Gaia}-DR3 star catalogue
 \item \textit{Gaia}-DR3 star catalogue
 \item Additional data for the \textit{Gaia}-DR3 star catalogue
\end{enumerate}


\section*{Software}

The following Tools and Python \texttt{packages} are used in this research:
\begin{center}
\begin{tabular}{lp{3.67cm}r}
\textbf{Package} & \textbf{Reference} & \textbf{Version}\\[0.15cm]

\texttt{galpy} & \citet{2015ApJS..216...29B} & \href{https://github.com/jobovy/galpy}{1.11.0.dev0} \\[0.125cm]

\texttt{scipy} & \citet{2020SciPy-NMeth} & \href{https://github.com/scipy/scipy}{1.16.0} \\[0.125cm]

Torus Mapper & \citet{2016MNRAS.456.1982B} & \href{https://github.com/PaulMcMillan-Astro/Torus}{-} \\[0.125cm]

\end{tabular}
\end{center}



\bibliographystyle{mnras}
\bibliography{bib/ref.bib}



\clearpage
\appendix


\section{Rotation matrices}\label{App1}

\subsection{Angle-Action coordinates}

The rotation matrix from the angle-action coordinates relative to the globular cluster ($\Delta\theta, \Delta J$) to the angle-action coordinates aligned with the principal axes of the stellar stream ($\Delta\bar{\theta}, \Delta\bar{J}$) is estimated from the \nbody\ simulation of the stream as described in Appendix F of \citetalias{2025MNRAS.539.2718P}. The obtained rotation matrix, which is used in this paper, is:
\begin{equation}
R \simeq \begin{pmatrix}
\phantom{-}0.689 & \s{3}-0.483 & \s{3}\phantom{-}0.540\\
-0.336 & \s{3}\phantom{-}0.448 & \s{3}\phantom{-}0.829\\
-0.642 & \s{3}-0.752 & \s{3}\phantom{-}0.147\\
\end{pmatrix}.
\end{equation}
We verify that this matrix is approximately equal to the rotation matrix obtained by diagonalising a linear approximation of the Hamiltonian function at the position of the globular cluster (\S 3, \citetalias{2025MNRAS.539.2718P}). We use the Torus Mapper algorithm (Section~\ref{actions_inv_trans}) with tolerance $\texttt{tol}=2\pd{-4}$ to numerically compute the Hessian matrix. By inverting the matrix whose columns are the eigenvectors of the Hessian matrix ordered in descending order of absolute eigenvalue magnitude, we obtain the following rotation matrix:
\begin{equation}
R_{\TM} \simeq \begin{pmatrix}
\phantom{-}0.693 & \s{3}-0.468 & \s{3}\phantom{-}0.549\\
-0.367 & \s{3}\phantom{-}0.426 & \s{3}\phantom{-}0.827\\
-0.620 & \s{3}-0.775 & \s{3}\phantom{-}0.124\\
\end{pmatrix}.
\end{equation}
We consider these two methods to be approximately equivalent, since the error in all values in $R_{\TM}$ is less than $20$ per cent of $R$.

\subsection{Sky coordinates}

The rotation from the ICRS sky coordinates ($\delta, \alpha$) to the sky coordinates aligned with the stream ($\phi_1, \phi_2$) is given by the following expression:
\begin{equation}\label{sky_R}
\begin{pmatrix}
\cos\svar{\phi_1} \cos\svar{\phi_2} \\
\sin\svar{\phi_1} \cos\svar{\phi_2} \\
\sin\svar{\phi_2}\\
\end{pmatrix}
= R_1 \times
\begin{pmatrix}
\cos\svar{\alpha} \cos\svar{\delta} \\
\sin\svar{\alpha} \cos\svar{\delta} \\
\sin\svar{\delta}\\
\end{pmatrix}.
\end{equation}
The rotation matrix $R_1$ is expressed in terms of the Euler angles $\alpha_x, \alpha_y \in \RangeNI{-\pi/2}{\pi/2}$ and $\alpha_z \in \RangeNI{0}{2\pi}$. It is estimated by minimising the following loss function:
\begin{equation}
 L_1\var{\alpha_x, \alpha_y} \equiv \sum \big(\phi_2 - \mean\svar{\phi_2}\big)^2,
\end{equation}
where the summation is defined for all stream stars in the GDR3 selection (Section~\ref{GDR3_selection}) and $\phi_2$ is computed using Eq.~\ref{sky_R} fixing the value of $\alpha_z=0$. The optimal values obtained for the angles are $(\alpha_x, \alpha_y)\simeq(-1.324, -0.316)$~rad. Then, a horizontal rotation of $\alpha_z=-3.440$~rad is included to position the closest extreme of the stream to the cluster at the origin of the coordinates. The resulting matrix is:
\begin{equation}
R_{1} \simeq \begin{pmatrix}
-0.997 & \s{3}-0.073 & \s{3}\phantom{-}0.025\\
-0.007 & \s{3}-0.237 & \s{3}-0.971\\
\phantom{-}0.077 & \s{3}-0.969 & \s{3}\phantom{-}0.236\\
\end{pmatrix}.
\end{equation}

Typically, the rotation is defined so that the average $\phi_2$ is approximately zero. This is achieved using the following loss function:
\begin{equation}\label{lk_2}
 L_2\var{\alpha_x, \alpha_y} \equiv \sum \phi_2^2.
\end{equation}
For the sake of completeness, we provide the rotation matrix corresponding to this definition, even though we only use $R_1$ in this paper. The optimal values obtained for the angles from Eq.~\ref{lk_2} and the horizontal rotation are $(\alpha_x, \alpha_y, \alpha_z)\simeq(-1.144, -0.524, -3.640)$~rad. The resulting matrix is:
\begin{equation}
R_{2} \simeq \begin{pmatrix}
-0.978 & \s{3}-0.198 & \s{3}\phantom{-}0.062\\
\phantom{-}0.014 & \s{3}-0.364 & \s{3}-0.931\\
\phantom{-}0.207 & \s{3}-0.910 & \s{3}\phantom{-}0.359\\
\end{pmatrix}.
\end{equation}


\section{Frequency distributions along the stream}\label{App2}

\begin{figure}
\includegraphics[width=1.0\columnwidth]{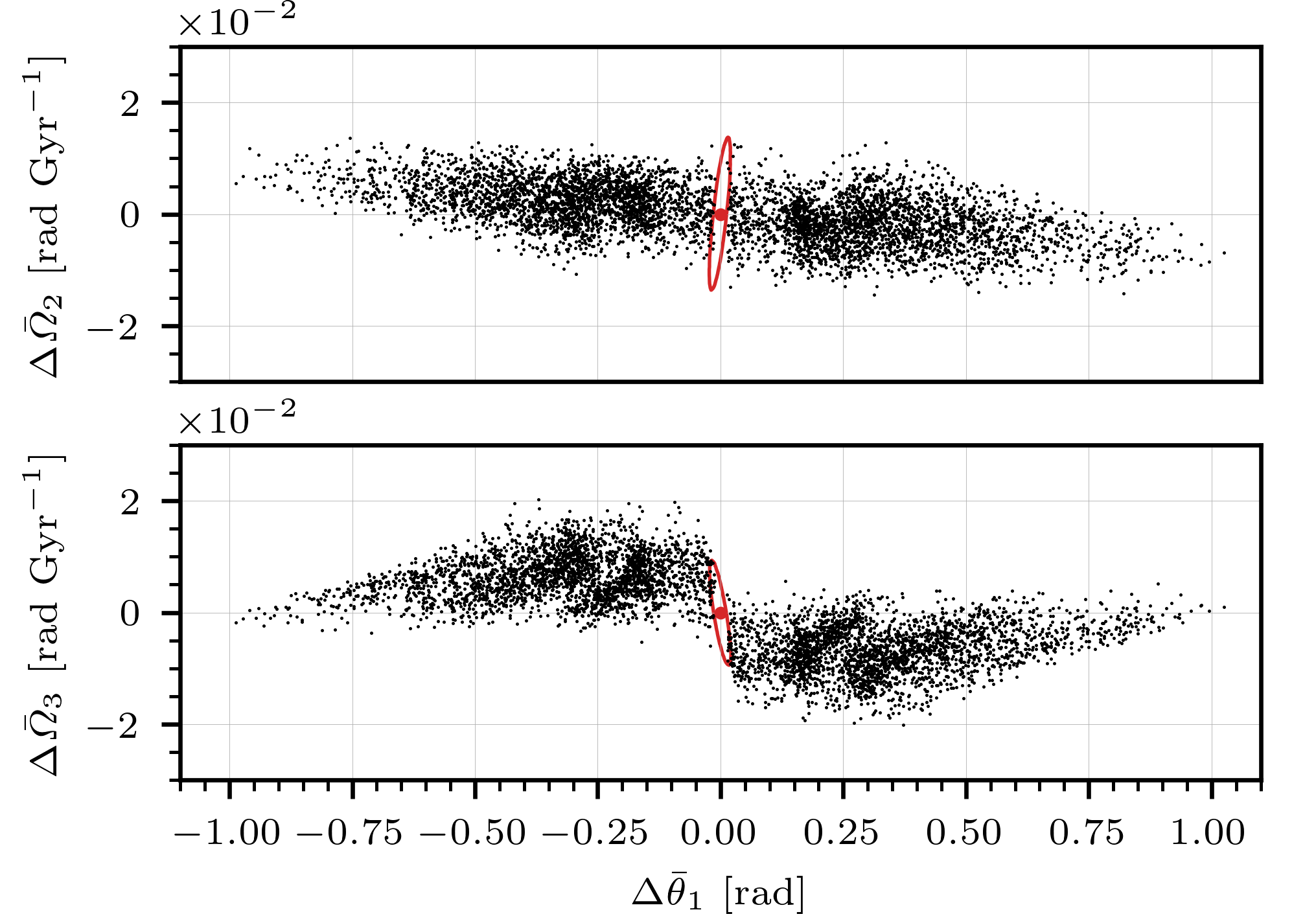}
\caption{Relative frequency perpendicular to the principal axis of the stream $\Delta\bar{\varOmega}_2$ (\textit{Top}) and $\Delta\bar{\varOmega}_3$ (\textit{Bottom}) as a function of the angle along the principal axis of the stream $\Delta\bar{\theta}_1$. The large red dot marks the position of the cluster and the red contour line the area containing 68 per cent of the cluster stars.}
\label{internal_structure}
\end{figure}

Figure~\ref{internal_structure} shows the frequencies of the simulated stream stars perpendicular to the stream as a function of the angle along the principal axis of the stream $\Delta\bar{\theta}_1$. The frequency space $\Delta\bar{\varOmega}_2$ (top panel) do not present a significant structure that allows us to identify the three internal components of the stream. Conversely, the internal components generated during the last two pericentre passages can be identified in the frequency space $\Delta\bar{\varOmega}_3$ (bottom panel). However, the component generated during the first pericentre passage is much more extended and populates the extremes of each arm. This makes it difficult to separate, as it is merged with the previous internal components.

\section{Double exponential function}\label{App3}

The double exponential function is defined in Eq.~\ref{double_exp} as a function of the radial angle of the progenitor cluster, corrected for the delay between the mass-loss peaks and the pericentre positions over the periodic domain $\theta_r^{\M} \in \RangeNI{0}{2\pi}$ (Section~\ref{num_stripp_stars}). This function depends on the parameters $A$ and $C$, as well as the scale lengths $\tau_1$ and $\tau_2$. Its integral over a general interval such that $0 \leqslant \theta_r^{\Inf} \leqslant \theta_r^{\Sup} < 2\pi$ has an analytical solution:
\begin{equation}\label{int_period}
\int_{\theta_r^{\Inf}}^{\theta_r^{\Sup}} \! f\var{\theta_r^{\M}}\,d\theta_r^{\M} = Ah\tau_1  + Ag\tau_2 \exp\!\left(-\frac{2\pi}{\tau_2}\right) + C(\theta_r^{\Sup} - \theta_r^{\Inf}),
\end{equation}
where $h$ and $g$ are defined as:
\begin{equation}
\begin{split}
h &\equiv \exp\!\left(-\frac{\theta_r^{\Inf}}{\tau_1}\right) - \exp\!\left(-\frac{\theta_r^{\Sup}}{\tau_1}\right) \\[0.45em]
g &\equiv \exp\!\left(\frac{\theta_r^{\Sup}}{\tau_2}\right) - \exp\!\left(\frac{\theta_r^{\Inf}}{\tau_2}\right).
\end{split}
\end{equation}

\subsection{Parameter estimation}\label{App3_par_est}

We use the \texttt{curve\_fit} function provided by the Python package \texttt{scipy} \citep{2020SciPy-NMeth} to compute the best-fitting parameters for a given two-dimensional data set. This function implements the Trust Region Reflective algorithm or \texttt{trf} method.

\subsection{Normalisation constant}\label{App3_norm_const}

For a general cluster orbit, defined from $t=-T$ to $t=0$, the angle $\theta_r^{\M} \in [\theta_r^{\T} \!\!\:,\:\! \theta_r^{\Z}]$, where $0 \leqslant \theta_r^{\Z} < 2\pi$. For this orbit, the inverse of the normalisation constant in Eq.~\ref{stripped_stars_pdf} is:
\begin{equation}
N_{\rm c}^{-1} = \int_{\theta_r^{\T}}^{\theta_r^{\Z}} f\var{\theta_r^{\M}} \,d\theta_r^{\M}.
\end{equation}
When $\theta_r^{\T} \geqslant 0$, this integral can be solved analytically using Eq.~\ref{int_period}. When $\theta_r^{\T} < 0$, the inverse of the normalisation constant can be expressed in terms of integrals within a single angular period. When expressing $\theta_r^{\T}$ in radians and unwrapping by a period of $2\pi$:
\begin{equation}
N_{\rm c}^{-1} = \int_{2\pi-\theta_r^{\R}}^{2\pi} f\var{\theta_r^{\M}} \,d\theta_r^{\M} + N_{\rm p} \! \int_{0}^{2\pi} \! f\var{\theta_r^{\M}}\,d\theta_r^{\M} + \int_{0}^{\theta_r^{\Z}} \! f\var{\theta_r^{\M}}\,d\theta_r^{\M},
\end{equation}
where $N_{\rm p}$ is the number of periods within $\theta_r^{\M} \in [\theta_r^{\T} \!\!\:,\:\! 0]$ and $\theta_r^{\R}$ is the remainder:
\begin{equation}
\begin{split}
N_{\rm p} &\equiv \big|\theta_r^{\T}\big| \,\modulus\, 2\pi \\[0.25em] \theta_r^{\R} &\equiv \big|\theta_r^{\T}\big| \;\!\reminder\;\! 2\pi.
\end{split}
\end{equation}

\subsection{Random sample}\label{App3_rdm_sample}

We generate random samples from the probability density function defined in Eq.~\ref{stripped_stars_pdf} using an inverse transform sampling algorithm. First, we unwrap the angle $\theta_r^{\M}$ by a period of $2\pi$ and evaluate $p\var{\theta_r^{\M}}$ at regular spacing within $\theta_r^{\M} \in [\theta_r^{\T} \!\!\:,\;\! \theta_r^{\Z}]$. We then compute the cumulative sum of these evaluations and normalise the result by subtracting the minimum value and dividing by the maximum. This ensures that the cumulative sum is contained within the interval $\Range{0}{1}$. The result obtained is a numerical evaluation of the cumulative distribution function (CDF). We compute the inverse of the CDF using the cubic spline method implemented by the \texttt{scipy} function \texttt{make\_interp\_spline}. The inverse is obtained when the parameter \texttt{x} is the CDF and \texttt{y} is the angle $\theta_r^{\M}$. Finally, we generate a random sample by evaluating the spline with a random sample from a uniform distribution defined on the interval $\Range{0}{1}$.


\section{Improved models}\label{App4}

We provide an improved model of the stripping points and frequency distribution along the principal axis of the stream using the left-skewed Gumbel distribution:
\begin{equation}
 G_{\rm L}\var{x\:\!|\:\!\mu,\sigma} \equiv \sigma^{-1} \exp\!\big(\:\!z - e^z \big),
\end{equation}
where $z\equiv (x-\mu)/\sigma$. This distribution provides better fits than a Gaussian distribution with the same number of free parameters.

\subsection{Angle \texorpdfstring{$\boldsymbol{\Delta\bar{\alpha}_1}$}{1} }

\begin{figure}
\includegraphics[width=1.0\columnwidth]{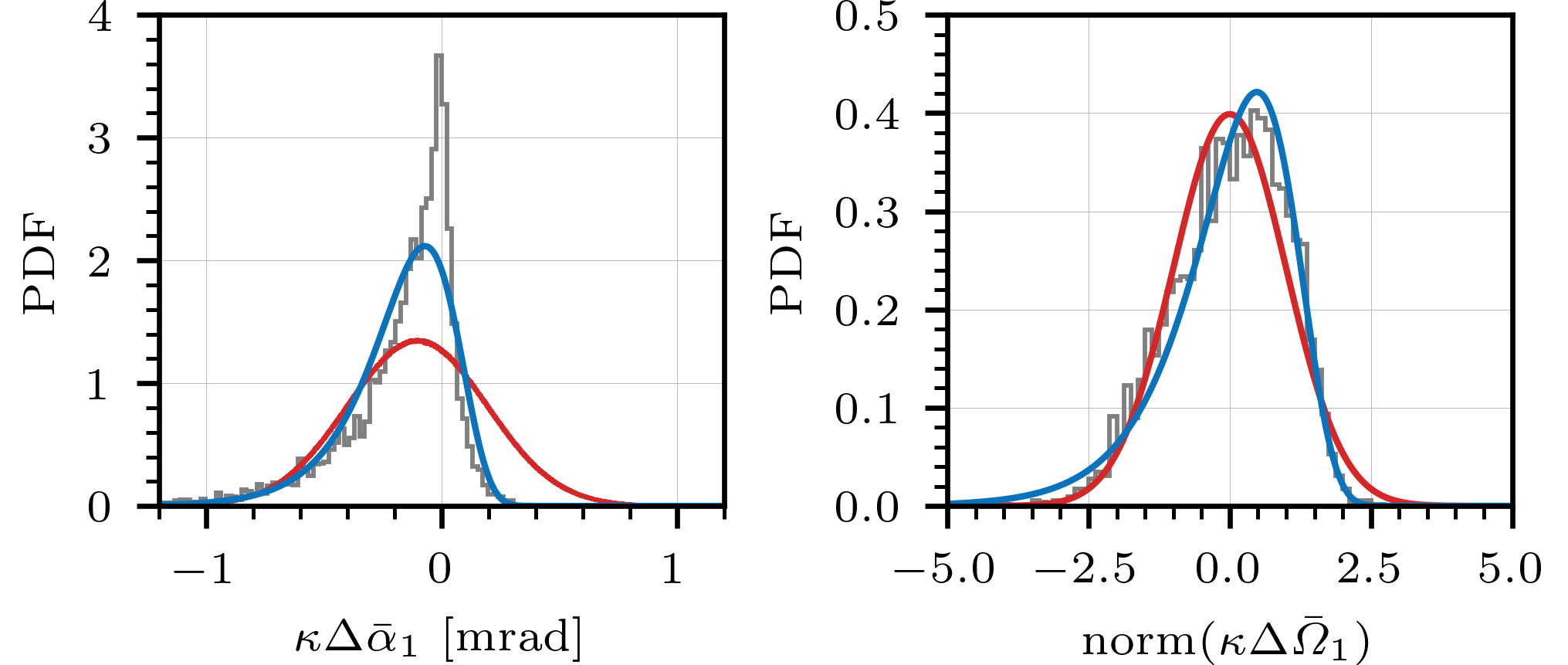}
\caption{\textit{Left:} Distribution of stripping points along the principal axis of the stream $\Delta\bar{\alpha}_1$, estimated using an histogram (grey line). First component of the multivariate Gaussian distribution introduced in Section~\ref{strip_dist} (red line). Best-fitting left-skewed Gumbel distribution (blue line). \textit{Right:} Distribution of the normalised frequency of the stripped stars along the principal axis of the stream $\Delta\bar{\varOmega}_1$ (grey line). Gaussian distribution centred at zero with a standard deviation of one (red line). Best-fitting left-skewed Gumbel distribution (blue line). In all panels, the symbol $\kappa$ indicates that both arms are shown together.}
\label{better_models}
\end{figure}

As shown by the grey histogram in the left panel of Figure~\ref{better_models}, the distribution of stripping points along the principal axis of the stream $\Delta\bar{\alpha}_1$ is skewed towards negative values. We also show the first component of the multivariate Gaussian distribution introduced in Section~\ref{strip_dist} as a red line. The distribution $G_{\rm L}\var{x\:\!|-0.072,\:\!0.174}$, plot in blue, provides a better fit than the Gaussian distribution, especially for positive values. Using distributions with more free parameters improves the fit to the peak region. For example, the Johnson SU and Generalised Logistic distributions provide better fits with four and three parameters, respectively.

\subsection{Frequency \texorpdfstring{$\boldsymbol{\Delta\bar{\varOmega}_1}$}{1} }

The frequency distribution of the stripped stars along the principal axis of the stream $\Delta\bar{\varOmega}_1$ is modelled using the mean and standard deviation of the simulated data within bins defined by the radial angle $\theta_r^{\M}$ (Section~\ref{freq_1_dist}). We normalise the values of $\Delta\bar{\varOmega}_1$  by subtracting the mean and dividing by the standard deviation within each bin as follows:
\begin{equation}
\normalised(\Delta\bar{\varOmega}_1) \equiv \frac{\Delta\bar{\varOmega}_1 - \mean(\Delta\bar{\varOmega}_1) }{ \std(\Delta\bar{\varOmega}_1) }.
\end{equation}
We then integrate all the data within the limits $\theta_r^{\M} \in \Range{-\pi}{\pi}$. The resulting distribution, estimated using a histogram, is plotted in grey in the right panel of Figure~\ref{better_models}. The best-fitting Gumbel distribution $G_{\rm L}\var{x\:\!|\:\!0.480,\:\!0.873}$ is plotted in blue, and a Gaussian distribution centred at zero with a standard deviation of one is plotted in red. The Gumbel distribution provides a slightly better fit because the data is left-skewed. Nevertheless, the Gaussian distribution is a good approximation for modelling the surface density of the stream.


\section{Pre-selection parameters for M68}\label{App5}

For the case of M68, the volume is constructed with $N=510$ Gaussian distributions computed using a bundle of $M=2\pd{4}$ orbits of length $l = 60$~Myr. The means of the Gaussian distributions used to compute the bundle of orbits are taken from Table~3 of \citetalias{2025MNRAS.539.2718P} for the position of the globular cluster, and from Table~2 of \citetalias{2025MNRAS.539.2718P} for the potential of the Milky Way. The scale factors of the Gaussian distributions are listed in Table \ref{pre_par}. Stars with an intersection $P_\SM{REG} \geqslant 6\pd{-6}$ ${\rm yr}^3 \, {\rm deg}^{-2} \, {\rm pc}^{-1} \, {\rm mas}^{-3}$ are pre-selected.

\begin{table}
\caption[]{\small{Scale factors of the Gaussian distributions used to compute the bundle of orbits for M68.}}

\begin{center}
\setlength{\tabcolsep}{3.5pt}
\begin{tabular}{lcccccc}
\toprule
&$r_{\rm h}$&$\delta$&$\alpha$&$v_r$&$\mu_\delta$&$\mu_{\alpha\ast}$\\
&\units{kpc}&\units{deg}&\units{deg}&\units{km s$^{-1}$}&\units{mas yr$^{-1}$}&\units{mas yr$^{-1}$}\\
\midrule
$\varepsilon_\mu$&0.25&2.5&2.5&0.2&0.05&0.05\\
\midrule
&$\rho_{\rm h}$&$a_{\rm h}$&$q_{h}$&$\beta_{\rm h}$&&\\
&\units{M$_{\odot}$ kpc$^{-3}$}&\units{kpc}&-&-&&\\
\midrule
$\varepsilon_\phi$&$10^6$&$2$&$0.25$&$0.1$&&\\
\bottomrule
\end{tabular}
\end{center}

\label{pre_par}
\end{table}


\section{ICRS surface density}\label{App6}

It is interesting to observe the peaks of the M68 stellar stream because we can use their location to estimate the time since the pericentre passages. This allows us to constrain the potential of the Milky Way by requiring the cluster to return to the pericentres after integrating backwards over the estimated time periods. To determine the location of the peaks and the expected number of stars, we simulate a stream with an age $T=3039.7$~Myr and a mass loss of $N_{T} \simeq 1.718$~stars~Myr$^{-1}$~arm$^{-1}$ using the method introduced in Section~\ref{model}.

\begin{figure}
\includegraphics[width=1.0\columnwidth]{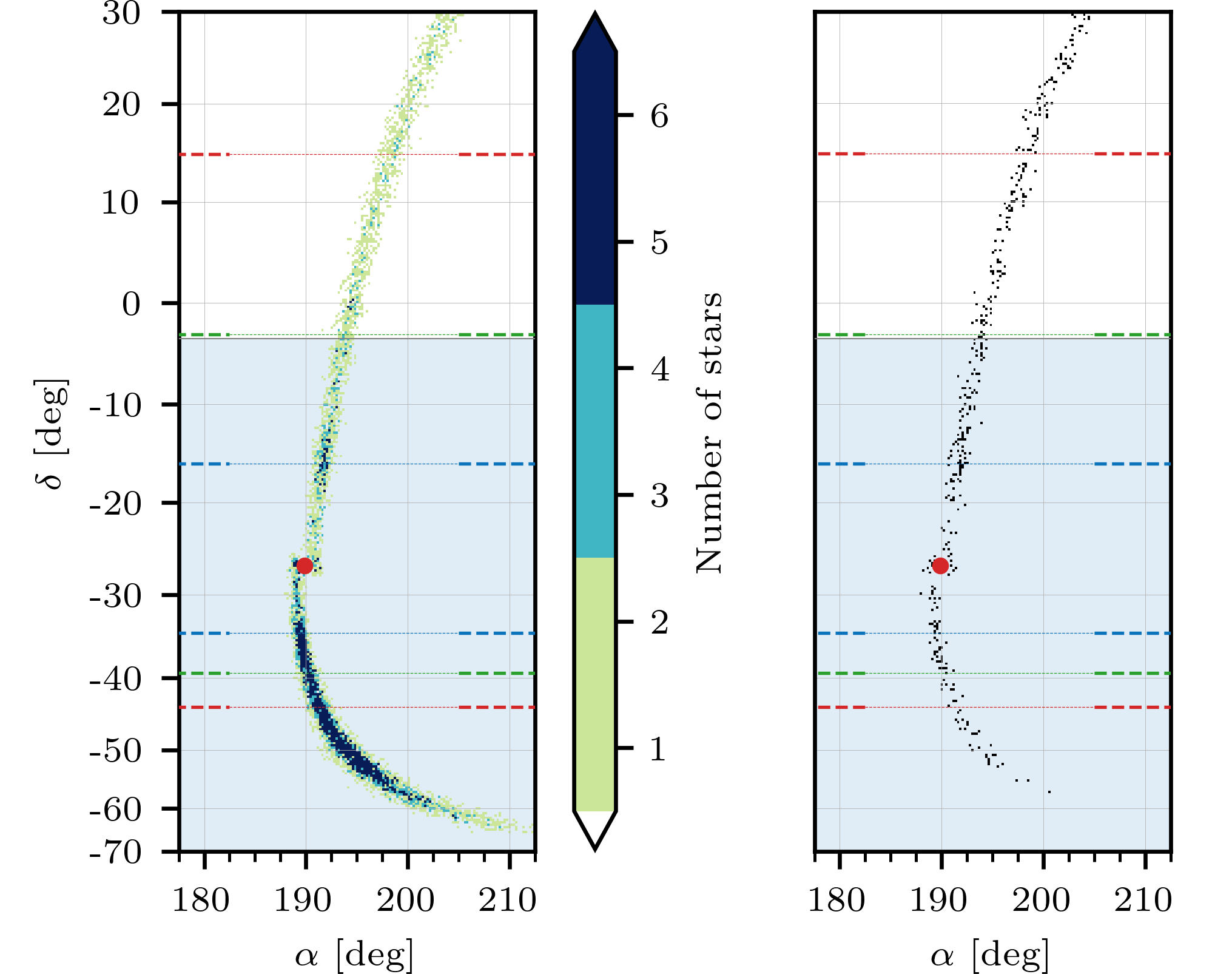}
\caption{Right ascension $\alpha$ and declination $\delta$ of a simulated tidal stream of age $T=3039.7$~Myr. The position of the cluster M68 is indicated by a large red dot. The position along the stream of the centre of the peaks corresponding to the first (red), second (green), and third (blue) pericentre passages are marked with coloured dashed lines. The area obscured by foreground star contamination is shaded in blue, and the limit is marked by a solid grey line ($\delta=-3.5$~deg). \textit{Left:} Projected surface density. The surface is divided into bins of equal area. \textit{Right:} Sample of simulated stars passing the \textit{Gaia}-DR3 selection function.}
\label{ICRS_density_projection}
\end{figure}

In the left panel of Figure~\ref{ICRS_density_projection}, we show the projected surface density in sky coordinates of the sample of $10\,442$ simulated stream stars. The declination has been scaled to divide the surface into equal square bins of area $16$~\textmu sr. We mark the position of the globular cluster M68 as a large red dot at $(\delta,\alpha)=(-26.744, 189.867)$~deg. The observable limit at $\delta=-3.5$~deg is marked by a solid grey line, and the area below this limit is shaded in blue. The stream is not visible in the blue area using only GDR3 data because it is obscured by foreground star contamination (Section~\ref{GDR3_selection}). In the right panel, we show a sample of $645$ stars that pass the \textit{Gaia} selection function (\S5.1, \citetalias{2025MNRAS.539.2718P}).

The coloured dashed lines indicate the positions along the stream of the peaks, following the colour code in Figure~\ref{m_loss} and \ref{angle_hist}. The peaks corresponding to the last pericentre passage (blue dashed lines) are clearly visible in the simulated data (left panel). Their location can be precisely determined because they mark the end of the underdensity between the cluster and the peaks in both arms. Due to projection distortion, the underdensity of each arm appears uneven in length. The exact location of the main peaks depends on the potential of the Milky Way, but generally, they always appear below the observational threshold.

Conversely, the second (green dashed line) and first (red dashed line) peaks of the leading arm lie within the observable area. However, as can be seen in the right panel of Figure~\ref{ICRS_density_projection}, the stream has a low surface density after the GDR3 selection function cut is applied. Additionally, the selection process and the presence of foreground contamination, neither of which is depicted in this figure, will complicate the identification of the peaks. Therefore, we conclude that identifying the peaks of the stream from its surface density using GDR3 data alone is problematic. However, it may be possible to identify the peaks if the signal-to-noise ratio of the star selection is improved. This could be achieved, for example, by incorporating radial velocity measurements into the selection methodology.


\section{Test fitting methodology}\label{App7}

\begin{figure}
\includegraphics[width=1.0\columnwidth]{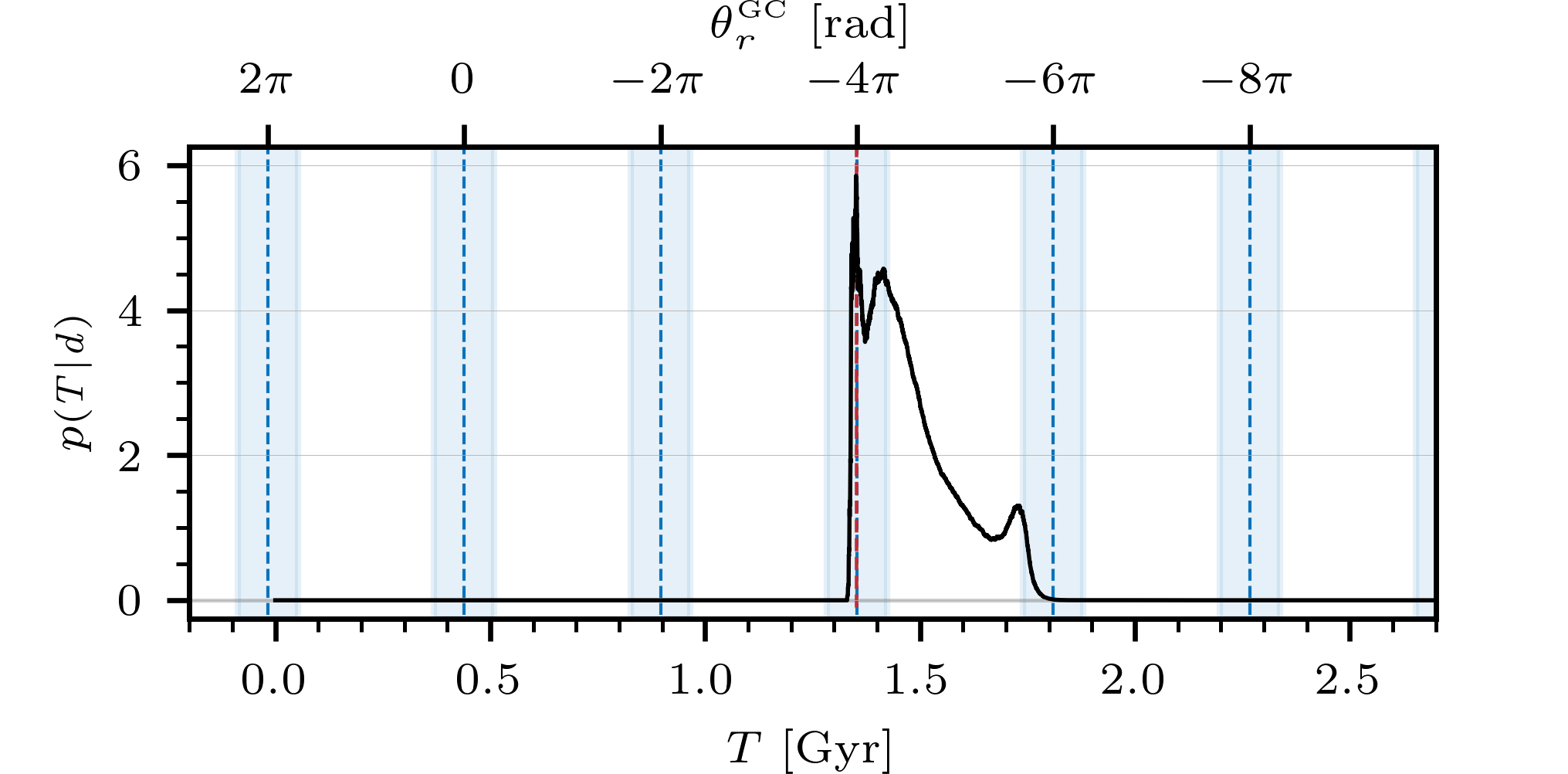}
\caption{Same as bottom panel of Figure~\ref{posterior}, but computed for a mock GDR3 star selection obtained from the \nbody\ simulation.}
\label{posterior_simulation}
\end{figure}

We test the methodology for constraining the age of the stellar stream and the mass loss of the globular cluster using a mock GDR3 star selection obtained from the \nbody\ simulation (\S5.4, \citetalias{2025MNRAS.539.2718P}). This sample consists of $116$ stars with magnitudes $G<20$~mag located in the main component of the stream. We compute the posterior distribution defined in Eq.~\ref{bayes} as described in Section~\ref{estimations} and plot the results in Figure~\ref{posterior_simulation}.

We observe that the posterior distribution peaks between approximately $T\in\Range{1.3}{1.8}$~Gyr, with a maximum at $\hat{T}=1351.3$~Myr. The value of the posterior distribution is approximately zero for streams with fewer than three pericentre passages. This is because the streams are too short. The maximum value approximately coincides with the first tidal shock experienced by the cluster at the pericentre at $t=-1353.28$~Myr, with a difference of $1.98$~Myr. For older streams, the posterior distribution decreases until the fourth pericentre passage. The relative maximums occur because the slower stars, released before and after the pericentre passages, populate the range $\Delta\bar{\theta}_1\Approx\Range{0.35}{0.45}$~rad of the observable section of the stream. This increases the surface density of the stream but not its length, thereby improving the fit. For streams with more than four pericentre passages, the posterior distribution is approximately zero, indicating that the streams are too long.

We estimate the age of the simulated stream to be $T=\round[2]{1.3513}_{-\round[2]{0.00889691}}^{+\round[2]{0.17122185}}$~Gyr using the methodology introduced in Appendix~\ref{App8} to determine asymmetric uncertainties. The relative uncertainties are of approximately $3$ and $12$ per cent of the estimated value, which we consider to be small. The number of stripped stars per unit of time per arm $N_T$ (Section~\ref{num_stripp_stars}) is determined as in Section~\ref{estimations}. We obtain $N_T \simeq 2.002\pm0.173$~stars~Myr$^{-1}$~arm$^{-1}$, which is approximately $10$ per cent smaller than the value of $2.23$~stars~Myr$^{-1}$~arm$^{-1}$ that best reproduces the \nbody\ simulation. We conclude that the discrepancies in determining the age and mass loss of the cluster are small, and that the methodology used is valid.


\section{Uncertainty determination}\label{App8}

Given a PDF $p\var{x}$ where $x\in(-\infty\,\text{,}\,\infty)$, and its maximum ${\hat{\mu}}$, we determine the upper limit of the $1\sigma$-level by solving the following equation for $\lambda_{\rm sup}\geqslant{\hat{\mu}}$:
\begin{equation}
 \int_{\hat{\mu}}^{\lambda_{\rm sup}} p\var{x} \, dx = \sigma \int_{\hat{\mu}}^\infty p\var{x} \, dx,
\end{equation}
where $\sigma\simeq0.683$. Similarly, we determine the lower limit by solving for $\lambda_{\rm inf}\leqslant{\hat{\mu}}$:
\begin{equation}
 \int_{\lambda_{\rm inf}}^{\hat{\mu}} p\var{x} \, dx = \sigma \int^{\hat{\mu}}_{-\infty} p\var{x} \, dx.
\end{equation}
The superior $1\sigma$-uncertainty of the variable $x$ is defined as $\epsilon_{\rm sup} \equiv \lambda_{\rm sup}-{\hat{\mu}}$, and the inferior as $\epsilon_{\rm inf} \equiv {\hat{\mu}}-\lambda_{\rm inf}$. The same procedure applies when $x$ is limited, except that the infinities are replaced by the limits.


\section{Age estimate using the entire GDR3 star selection}\label{App9}

The best-fitting age of the M68 stream estimated using only the stars in the main component is $\hat{T}=3039.7$~Myr (Section~\ref{estimations}).This measurement is compared to the estimate using all the stars in the GDR3 selection, which includes both the main and the envelope components. In this case, we obtain $\hat{T}=2637.4$~Myr. This difference is due to the fact that the stars in the envelope populate the region $\Delta\bar{\theta}_1 \LessSim 0.8$~rad of the observable section of the stream. Therefore, a stream with a larger surface density close to the cluster is favoured. This is achieved by a younger stream that has undergone one fewer pericentre passage. Using the method of Appendix~\ref{App8}, we estimate the errors, obtaining $T=\round[2]{2.6374}_{-\round[2]{0.1483826}}^{+\round[2]{0.25761837}}$~Gyr. This result is closer to the $T=2.34\pm0.82$~Myr obtained by \citet{2025ApJ...980L..18C}, indicating that the difference is partially caused by the exclusion of the stars in the envelope of the stream.



\section{Constant mass loss and frequency distribution}\label{App10}

We eliminate the angular dependency of the double exponential model (Eq.~\ref{double_exp}) by setting the amplitude $A=0$. In this case, the model is reduced to $f\var{\theta_r^{\M}} = C$. This implies that the mass loss of the cluster is constant, and that the frequency distribution of the stripped stars along the principal axis of the stream $\Delta\bar{\varOmega}_1$ is Gaussian (Section~\ref{freq_1_dist}). The values that best reproduce the results of the \nbody\ simulation are given by: $N_{\rm s}\simeq29.371$ ${\rm stars}\times(4\pi/139\,{\rm rad})^{-1}$, $\mean(\Delta\bar{\varOmega}_1)=0.391$~mrad~Myr$^{-1}$, and $\std(\Delta\bar{\varOmega}_1)=0.115$~mrad~Myr$^{-1}$. Hereafter, we refer to this as the `Gaussian Model', and to the model introduced in Section~\ref{model} as the `Variable Model'.

\begin{figure}
\includegraphics[width=1.0\columnwidth]{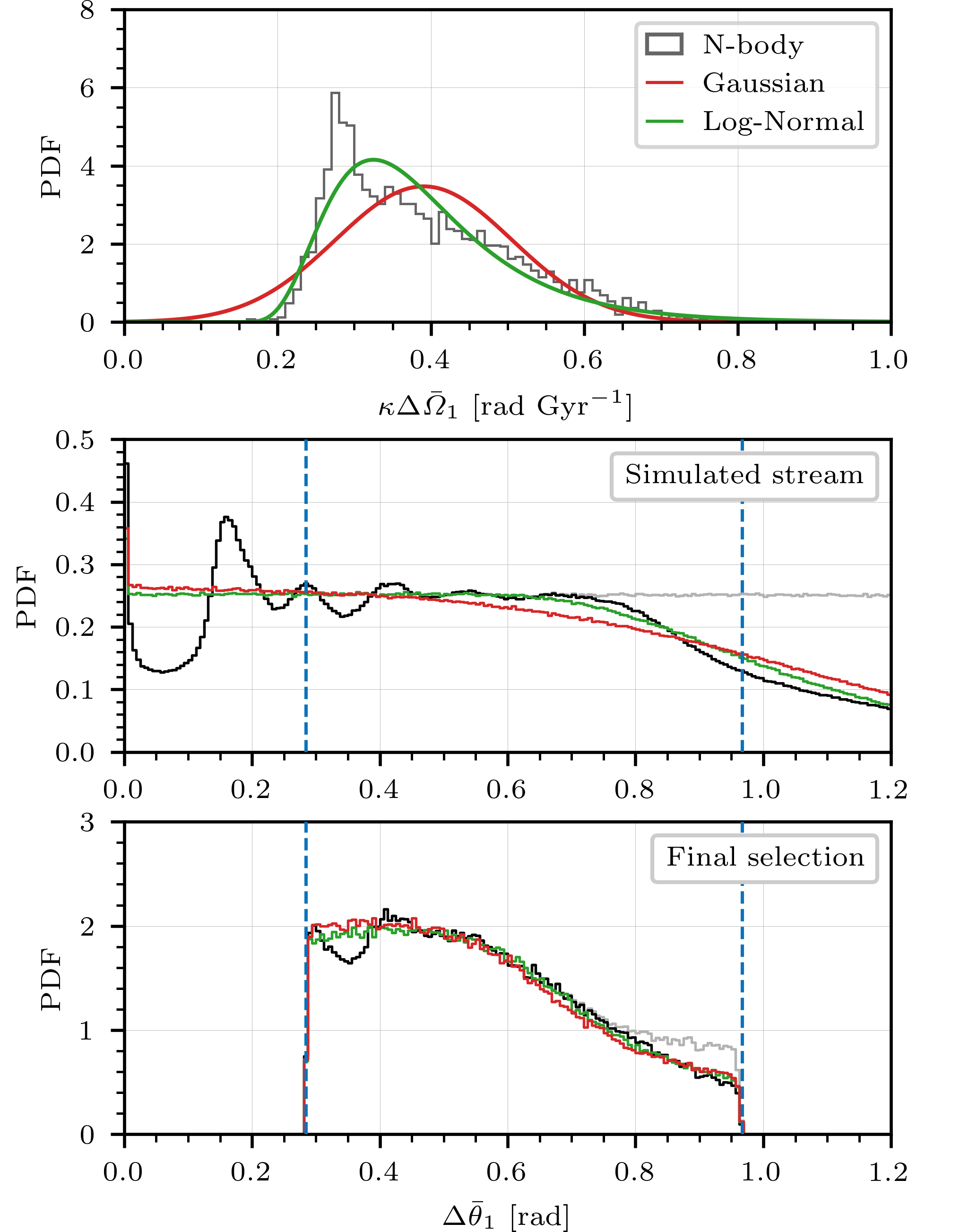}
\caption{\textit{Top:} Distribution of frequencies of the stripped stars along the principal axis of the stream $\Delta\bar{\varOmega}_1$, estimated with an histogram (grey). The best-fitting Gaussian distribution is shown in red and the best-fitting Log-Normal distribution in green. The symbol $\kappa$ indicates that both arms are shown together. \textit{Middle and Bottom:} Same as the top and middle-bottom panels of Figure~\ref{surface_density_model}. The red and green histograms show the PDF of the surface density of a simulated stream of age $T=3039.7$~Myr, computed assuming a constant mass loss for the globular cluster and constant frequency distribution for the stream stars. The red histogram shows the case where $\Delta\bar{\varOmega}_1$ is assumed to be Gaussian and the green histogram shows the case where it is assumed to be Log-Normal.}
\label{const_mass_loss}
\end{figure}

The top panel of Figure~\ref{const_mass_loss} shows the distribution of $\Delta\bar{\varOmega}_1$ of the stream stars from the \nbody\ simulation as a grey histogram. We estimate this distribution using only the stars stripped during the two complete periods of the simulation, which correspond to $\theta_r^{\M}\in\RangeNI{-4\pi}{0}$. This distribution has a sharp peak at $\Delta\bar{\varOmega}_1 \in \Range{0.25}{0.3}$~rad~Gyr$^{-1}$, corresponding to the slow stars stripped around the apocentres. The stars stripped at the pericentres move faster and populate the tail of the distribution. The best-fitting Gaussian distribution is shown in red. This model cannot reproduce the skewness of the distribution or its peak. However, the Gaussian distribution accurately reproduces the far upper tail of the distribution, which ensures that a stream simulated with the Gaussian Model will be of a similar length to that obtained with the \nbody\ simulation.

In the middle panel of Figure~\ref{const_mass_loss}, we show the PDF of the surface density of a simulated stream as a function of the angle $\Delta\bar{\theta}_1$. As in Figure~\ref{surface_density_model}, the black histogram corresponds to a stream of age $T=3039.7$~Myr, and the grey histogram corresponds to a saturated stream over the entire angular range depicted. Both streams are simulated using the Variable Model. The red histogram shows the surface density of a simulated stream of age $T=3039.7$~Myr computed using the Gaussian Model. This model cannot reproduce the underdensity next to the globular cluster ($\Delta\bar{\theta}_1=0$) or the peaks. In addition, it does not produce a constant saturation density, but a curvilinear surface density that underestimates $\Delta\bar{\theta}_1\Approx\Range{0.3}{0.9}$~rad and overestimates $\Delta\bar{\theta}_1\GtrSim0.9$~rad with respect to the Variable Model (black histogram).

In the bottom panel of Figure~\ref{const_mass_loss}, we plot the distribution of stars from the main component of the simulated stream that pass the GDR3 selection function and the selection process (Section~\ref{mock_selection}). The Gaussian Model (red histogram) presents a flat PDF within the range $\Delta\bar{\theta}_1\Approx\Range{0.3}{0.5}$~rad, which appears incompatible with the observational data shown in Figure~\ref{surface_density}. We test whether this model can accurately determine the accretion time of the simulated globular cluster by using a mock star sample obtained from the \nbody\ simulation, as described in Appendix~\ref{App7}. We obtain an age of $T=1.51\pm0.08$~Gyr, which overestimates the time since the first pericentre passage by approximately $159$~Myr.

The model with constant mass loss can be improved assuming that the frequency distribution of the stream stars along the principal axis of the stream $\Delta\bar{\varOmega}_1$ is a Log-Normal distribution of equation:
\begin{equation}
\logN\!\var{x|\mu,\sigma,s} \equiv \frac{1}{\sqrt{2\pi}\!\:(x-\mu)\!\:s} \exp\!\left[\frac{-\log\!\left(\frac{x-\mu}{\sigma}\right)^2}{2s^2} \right],
\end{equation}
with parameters $\mu\simeq0.125$~mrad~Myr$^{-1}$, $\sigma\simeq0.242$ mrad~Myr$^{-1}$, and $s\simeq0.436$. We plot this distribution as a green line in the top panel of Figure~\ref{const_mass_loss}. Although the Log-Normal distribution accurately describes the skewness of the distribution, it does not reproduce the peak corresponding to the slow stars stripped at the apocentre. Furthermore, the Log-Normal distribution overestimates the fraction of stars with $\Delta\bar{\varOmega}_1\GtrSim0.7$~rad~Gyr$^{-1}$, resulting in simulated streams that are longer than those obtained with the \nbody\ simulation.

The middle panel of Figure~\ref{const_mass_loss} shows as a green histogram the surface density of a simulated stream of age $T=3039.7$~Myr computed assuming a constant mass loss and assuming that $\Delta\bar{\varOmega}_1$ follows the Log-Normal distribution. We refer to this model as the `Log-Normal Model'. As in the Gaussian case, this model does not reproduce the density variations close to the cluster or the density distribution slope around $\Delta\bar{\theta}_1\Approx1$~rad obtained with the Variable Model (black histogram). However, it reproduces the constant density saturation level predicted by the Variable Model. In the bottom panel of Figure~\ref{const_mass_loss}, we plot the distribution of the final star selection obtained using the Log-Normal Model as a green histogram. This distribution is similar to that obtained using the Gaussian Model. Therefore, the estimated age of the stream is also similar, being $T=1.60^{+0.08}_{-0.11}$~Myr, which overestimates the age of the \nbody\ stream by approximately $246$~Myr.

We conclude that the models assuming constant mass loss provide a good approximation of the stream surface density far from the progenitor cluster, where the density peaks are negligible. While the Gaussian Model can reproduce the length of the stream, it cannot reproduce the constant saturation density. Furthermore, it overestimates the age of the \nbody\ simulation of the M68 stream due to the inaccurate modelling of the density peaks. On the other hand, the Log-Normal Model requires an additional free parameter compared to the Gaussian Model, but it can accurately reproduce the stream saturation level. However, as with the Gaussian Model, it overestimates the age of the stream. In addition, it overestimates the length of the stream by a factor of approximately two. Therefore, in order to accurately estimate the age of the M68 stream, we consider it necessary to model the density variations caused by variable mass loss and variable frequency distribution of the stream stars.



\bsp	
\label{lastpage}

\end{document}